\documentclass[notitlepage,prd,twocolumn,showpacs,preprintnumbers,nofootinbib,superscriptaddress,floatfix,tightenlines]{revtex4-1}
\usepackage{mathrsfs}
\usepackage{amssymb}
\usepackage{amsmath}
\usepackage{graphicx}
\usepackage{amsfonts,amsbsy}
\usepackage{nicefrac}
\usepackage{xcolor}
\definecolor{lcolor}{rgb}{0.5,0,0}
\definecolor{citcolor}{rgb}{0,0.3,0.0}
\usepackage[breaklinks,colorlinks,urlcolor=blue,citecolor=citcolor,linkcolor=lcolor]{hyperref}

\newcommand{\mbf}{\mathbf}

\newcommand{\tpert}{t_{\text{pert}}}
\newcommand{\tcent}{\bar{t}}			
\newcommand{\mrm}{\mathrm}
\newcommand{\ah}{\mrm{ah}}
\newcommand{\Tr}{\mrm{Tr}}
\newcommand{\HTL}{\mrm{HTL}}
\newcommand{\fit}{\mrm{fit}}
\newcommand{\rel}{\mrm{rel}}
\newcommand{\qs}{Q_\mathrm{s}}
\newcommand{\pInit}{p_0}
\newcommand{\Q}{Q}
\newcommand{\wplas}{\omega_{\mrm{pl}}}
\newcommand{\fig}{Fig.~}
\newcommand{\figs}{Figs.~}
\newcommand{\eq}{Eq.~}
\newcommand{\se}{Sec.~}

\newcommand{\nr}[1]{(\ref{#1})}

\newcommand{\ud}{\mathrm{d}}
\newcommand{\ii}{{\boldsymbol{\hat{\i}}}}
\newcommand{\jj}{{\boldsymbol{\hat{\j}}}}
\newcommand{\kk}{{\mathbf{\hat{k}}}}
\newcommand{\cl}{\mrm{cl}}
\newcommand{\uj}{\mrm{j}}
\newcommand{\dtmax}{\Delta t_{\text{max}}}

\begin{document}

\title{Spectral function for overoccupied gluodynamics from real-time lattice simulations}

\author{K.~Boguslavski} 
\affiliation{Department of Physics, P.O.~Box 35, 40014 University of Jyv\"{a}skyl\"{a}, Finland}

\author{A.~Kurkela} 
\affiliation{Theoretical Physics Department, CERN, Geneva, Switzerland}
\affiliation{Faculty of Science and Technology, University of Stavanger, 4036 Stavanger, Norway}

\author{T.~Lappi} 
\affiliation{Department of Physics, P.O.~Box 35, 40014 University of Jyv\"{a}skyl\"{a}, Finland}
\affiliation{Helsinki Institute of Physics, P.O.~Box 64, 00014 University of Helsinki, Finland}

\author{J.~Peuron} 
\affiliation{Department of Physics, P.O.~Box 35, 40014 University of Jyv\"{a}skyl\"{a}, Finland}

\begin{abstract}
We study the spectral properties of a highly occupied non-Abelian non-equilibrium plasma appearing ubiquitously in weak coupling descriptions of QCD matter. The spectral function of this far-from-equilibrium plasma is measured by employing linear response theory in classical-statistical real-time lattice Yang-Mills simulations. We establish the existence of transversely and longitudinally polarized quasiparticles and obtain their dispersion relations, effective mass, plasmon frequency, damping rate and further structures in the spectral and statistical functions. Our new method can be interpreted as a non-perturbative generalization of hard thermal loop (HTL) effective theory. We see indications that our results approach leading order HTL in the appropriate limit. The method can also be employed beyond the range of validity of HTL.
\end{abstract}

\maketitle


\section{Introduction}
Strong Yang-Mills fields are a ubiquitous feature of weak coupling descriptions of Quark-Gluon Plasma and heavy ion collisions. They are created through different mechanisms and affect a broad range of phenomena. 
For instance, in thermal equilibrium the infrared tail of thermal Bose distribution $f\sim T/\omega$ makes
low frequency modes highly occupied, corresponding to strong, long-wavelength fields. The interaction of these fields 
with hard modes at the scale $T$ becomes nonperturbative at the soft or asymptotic mass scale $ m  \sim (\alpha_s\int \ud^3 p\, f/p)^{1/2} \sim g T$. The physics of these soft modes affects the weak coupling expansion of the equation of state starting from the $\mathcal{O}(g^3)$ level \cite{Kapusta:1979fh,Blaizot:1999ap}.
These same modes contribute to the thermal photon and low-mass dilepton production rates already at leading
order (LO) because of the infrared sensitivity of the $t$-channel exchange \cite{Kapusta:1991qp}. Similarly the effective kinetic theory
description used to describe near-equilibrium transport in weakly coupled QCD is sensitive to the physics
of the soft modes already at LO \cite{Arnold:2002zm}. 

Strong classical Yang-Mills fields also appear in the weak coupling description of the initial stages
of heavy-ion collisions. The initial condition of the post-collision
debris in the midrapidity region is dominated by nonperturbatively strong gluon fields at the characteristic
momentum scale $\qs$ with mode-occupancies  $f\sim 1/g^2$ \cite{Lappi:2006fp,Gelis:2015gza}. Once the dominant modes of these fields have diluted because of the combined effect of expansion and non-perturbative self-interactions, the subsequent approach towards local thermal equilibrium is described by the effective kinetic theory that is, again, sensitive to strong infrared fields \cite{Kurkela:2015qoa}. 

In most of the above mentioned cases, there is a scale separation between the highly occupied soft ($p\sim  m $) modes, and the 
hard ($p \sim \Lambda$) modes with which they interact nonperturbatively. In this case, the real-time dynamics of the soft 
modes can be followed in the Hard (Thermal) Loop (HTL) effective theory \cite{Braaten:1989mz,Blaizot:2001nr} in which the soft classical fields are coupled to hard ballistically propagating point-like color charges that interact with the soft classical fields through colored Vlasov-Wong equations \cite{Blaizot:2001nr}. The Hard Loop dynamics can then be solved perturbatively in the self-interactions of the soft modes (while still treating 
the interaction with the hard particles nonperturbatively, i.e., performing the HTL resummation). The small parameter in this
expansion is given by the scale separation $ m / \Lambda$, which in thermal equilibrium is  $gT/T$.
 This expansion parameter can be significantly larger 
than that of the hard sector ($\alpha_s/4\pi$), and the extent to which the weak coupling expansions of the various quantities are reliable is dependent on how well the soft sector is described by the LO HTL expressions. 

Computing complicated observables and higher order corrections in HTL theory is technically involved and is a major 
limiting factor in the advancement of the program of thermal weak coupling calculations \cite{CaronHuot:2007gq,CaronHuot:2008uh}. For space-like correlation functions 
powerful Euclidean techniques exist \cite{Appelquist:1981vg,CaronHuot:2008ni}. These techniques have made high order calculations of the equation of state at finite temperature possible \cite{Kajantie:2002wa}, and even allowed for a numerical evaluation of the full contributions arising from the soft sector through the simulation of an effective theory \cite{Hietanen:2004ew}. They have also enabled several next-to-leading order (NLO) computations, including the  photon and small-mass dilepton production rates  \cite{Ghiglieri:2013gia,Ghiglieri:2014kma} among other results.

Lattice formulations of HTL in real time (see e.g.~\cite{Hu:1996sf,Moore:1997sn,Tang:1997gk,Bodeker:1999gx,Dumitru:2007rp,Schenke:2008gg,Rebhan:2008uj,Attems:2012js}) rely on an explicitly separate description of the soft modes as gauge fields and of the hard modes as classical particles. There are two possible approaches used in the literature. One is to initialize the fields with a (classical) thermal distribution, where most of the field energy resides in the modes close to the lattice UV cutoff $1/a_s$. In this case the inability to renormalize this system prevents following the time evolution of the soft fields in a controlled way and prohibits a numerical evaluation 
of time-like correlation functions (see however \cite{Schenke:2008gg}). The other option used e.g. in studies of plasma instabilities~\cite{Dumitru:2005gp,Nara:2005fr,Bodeker:2007fw} is to initialize the classical field with a sufficiently UV safe distribution. In this latter case, in order to correctly resolve the soft physics free of cutoff artefacts, one needs $ m  \ll 1/a_s$. On the other hand one needs to simultaneously have $\Lambda \gg 1/a_s$ in order for the hard modes to be sufficiently  localized to justify their description as classical particles.  These requirements make the simulations rather expensive. More importantly they make it impossible to study systems with a smaller scale separation, i.e. larger values of $ m / \Lambda$, and thus inherently difficult to extend the HTL theory beyond strict leading order. We would like to argue here that a formulation in terms of classical fields and linearized fluctuations provides a method to describe a similar physical system in a way which can also be used in a regime where the scale separation $ m / \Lambda$ need not be small. This enables one to go beyond the strict leading order in $ m / \Lambda$ also in nonperturbative lattice calculations for time-like correlation functions.

The aim of this paper is twofold. First, we study a specific isotropic and over-occupied far-from-equilibrium system that, starting from an overoccupied initial condition, is undergoing a cascade of energy towards the UV in a self-similar regime~\cite{Berges:2008mr,Kurkela:2011ti,Kurkela:2012hp,Berges:2012ev,Schlichting:2012es,Berges:2013fga,York:2014wja}. This system exhibits an increasing separation of scales between a hard scale $\Lambda$ that dominates the energy density, and the softer scale $\sim  m $ where the interaction with the hard scale becomes non-perturbative. This system is similar to thermal equilibrium at weak coupling in the sense that there is a scale separation $ m  \ll \Lambda$ (or $gT\ll T$). On the other hand, it differs from a thermal system in the sense that this scale separation is not given by the coupling, but rather increases with time as $ m / \Lambda \sim (\Q t)^{-2/7}$, where $\Q$ is a momentum scale constant in time. In this system, also the hard sector is over-occupied $f(p \sim \Lambda) \gg 1$ with $\Lambda \ll 1/a_s$ and hence, the time evolution can be followed numerically within a classical Yang-Mills simulation \cite{Aarts:2001yn,Mueller:2002gd,Jeon:2004dh,Skullerud:2003ki,Arrizabalaga:2004iw,Berges:2004yj,Kurkela:2012hp}.  We compare the expectations of LO HTL theory to the full numerical time evolution. Doing so, we confirm and quantify to what extent HTL at LO is a good approximation to describe the physics of soft modes of this system. Similar comparisons have been done for scalar theory~\cite{Aarts:1997kp,Aarts:2001yx,Epelbaum:2011pc,Schachner:2016frd,Orioli:2018xxx}, while for gauge theory previous studies of the quasiparticle dispersion relation~\cite{Krasnitz:2000gz,Berges:2012ev,Lappi:2016ato,Lappi:2017ckt} have only used the behavior of the background field (i.e. the statistical function) at equal times. 

Secondly, we turn the argument around and interpret the classical simulation as a non-perturbative simulation of HTL. 
As HTL theory is insensitive to the detailed form of the hard sector, all isotropic equilibrium and non-equilibrium systems exhibiting
the large scale separation are related within HTL theory. Therefore the results obtained from the controlled classical simulation of classical fields
can be directly applied to thermal equilibrium where no other controlled non-perturbative methods exist.\footnote{Methods based on reconstruction of gauge-fixed gluonic spectral functions from Euclidean lattice data (e.g., \cite{Jarrell:1996rrw}) are usually applied to low or moderate temperatures and typically require further input like perturbative or analytical insight to obtain a spectral function within reasonable error bars, where recent advances have significantly improved the results (e.g., \cite{Haas:2013hpa,Dudal:2013yva,Strauss:2012dg,Cyrol:2018xeq}). Our method on the other hand does not require prior input or spectral reconstruction and corresponds to directly probing the spectral function at high temperatures and weak couplings, non-perturbatively extending the HTL formalism.}
As an application we determine the plasmon damping rate generalizing the classic result from Braaten and Pisarski \cite{Braaten:1990it} to finite momentum. 

We will start in \se \ref{sec:methods} by introducing the considered theory, relevant correlation functions, predictions from the HTL theory and our numerical method. Our results are shown in \se \ref{sec:results}. We will conclude in \se \ref{sec:conclu}.


\section{Theory}
\label{sec:methods}

\subsection{Classical Non-Abelian gauge theory}
\label{sec_class_nonAb_theory}

We consider an $\mrm{SU}(N_c)$ gauge theory~\footnote{Based on other related studies~\cite{Berges:2008zt,Ipp:2010uy,Berges:2017igc} we do not expect any qualitative difference between $N_c = 2$ and $N_c = 3$.} with $N_c = 2$ in temporal $A_0 = 0$ gauge.
The classical equations of motion in continuum are given by
\begin{align}
 \label{eq_EOM_BG_cont}
 \partial_t A_i^a(t,\mbf x) \,&= E^{i,a}(t,\mbf x) \nonumber \\
 \partial_t E^{i,a}(t,\mbf x) \,&= D^{ab}_j(t,\mbf x) F^{b,ji}(t,\mbf x),
\end{align}
with the gauge field $A_i$ and the (chromo-)electric field $E^i$, (spatial) vector components $i,j = 1,2,3$ and the adjoint group indices $a,b = 1,\dots,N_c^2-1$. The covariant derivative and the field strength tensor are given by $D^{ab}_j = \delta^{ab}\partial_j - g f^{abc} A_j^c$ and $F_{ij}^a = \partial_i A_j^a - \partial_j A_i^a + i g f^{abc}A_i^b A_j^c$, respectively. Here $g$ is the coupling constant and the structure constants $f^{abc}$ for $\mrm{SU}(2)$ are given by the totally anti-symmetric Levi-Civita tensor $f^{abc} = \epsilon^{abc}$.

The fields are discretized on a cubic lattice with $N_s$ lattice sites with lattice spacing $a_s$ in each of the $3$ dimensions. In order to preserve gauge invariance, one uses link variables $U_j(t, \mbf x)$, that are related to the gauge fields by $U_j(t, \mbf x) \approx \exp\left( ig a_s A_j^a(t,\mbf x)\Gamma^a \right)$, where $\Gamma^a$ are the generators of the $\mrm{su}(N_c)$ algebra normalized in the standard way, i.e. $\mathrm{Tr}\left(\Gamma^a \Gamma^b\right) = \nicefrac{1}{2}\, \delta^{ab}.$ With $E^j = E^j_a \Gamma_a$, the time evolution then follows from the discretized equations of motion 
\begin{align}
\label{eq_EOM_BG}
 U_j(t + \ud t/2, \mbf x) \,&= e^{i \ud t\, a_s g E^j(t,\mbf x)} U_j(t - \ud t/2, \mbf x) \nonumber \\
 g E^i(t + \ud  t,\mbf x) \,&= g E^i(t,\mbf x) \nonumber \\
 -\frac{\ud t}{a_s^3} \sum_{j \neq i} \bigg[U_{ij}&\Big(t - \frac{\ud t}{2}, \mbf x\Big) + U_{i(-j)}\Big(t - \frac{\ud t}{2}, \mbf x\Big)\bigg]_{\ah},
\end{align}
where the plaquette is given by $U_{ij}(\mbf x) = U_i(\mbf x) U_j(\mbf x + \ii) U_i^\dagger(\mbf x + \jj) U_j^\dagger(\mbf x)$ while the plaquette in the negative $j$ direction is $U_{i(-j)}(\mbf x) = U_i(\mbf x) U_j^\dagger(\mbf x + \ii - \jj) U_i^\dagger(\mbf x - \jj) U_j(\mbf x - \jj)$, with $\ii$ and $\jj$  unit vectors in the $i,j$ directions. The antihermitian traceless part of a matrix $V$ is defined as 
\begin{align}
 \left[V\right]_\ah \equiv \frac{-i}{2} \left( V- V^\dagger - \frac{1}{N_c} \Tr\left(V - V^\dagger\right)\right). 
\end{align}
The discretized equations of motion preserve the Gauss law condition at every time step
\begin{align}\label{eq:gauss}
 D^{ab}_j\, E^{b,j}(t, \mbf x) = 0,
\end{align}
where the discretized covariant (backward) derivative is (suppressing the time variable for brevity) given by $D^{ab}_j\, V^b(\mbf x) = (V^a(\mbf x) - U_j^{\dagger,ab}(\mbf x - \jj) V^b(\mbf x - \jj)) / a_s$. The parallel transporting link field in adjoint representation reads $U_j^{ab}(\mbf x) \equiv 2\Tr\left(\Gamma^a\, U_j(\mbf x)\, \Gamma^b\, U_j^\dagger(\mbf x) \right)$. 

To work in Fourier space, we use the common definition of the spatial Fourier transform of a field $w$ as $w(\mbf x) = \frac{1}{V} \sum_{\mbf p} e^{i \mbf x \cdot \mbf p} w(\mbf p)$, with the volume $V = a_s^3 N_s^3$. The discrete momenta are given by $p_j = 2\pi j /(a_s N_s)$. On the other hand, the backward derivative is defined as $-i\partial^B_j w(\mbf x) = -i/a_s \left(w(\mbf x) - w(\mbf x - \jj)\right)$. We define the (complex-valued) backward momentum on the lattice as its eigenvalue
\begin{align}
 p^B_j = \frac{-i}{a_s} \left(1 - e^{-i a_s p_j}\right). 
\end{align}
The forward derivative leads to the momentum $p^F_j = (p^B_j)^*$ and the discretized second derivative $-\partial^2_j = -\partial^B_j\partial^F_j = -\partial^F_j\partial^B_j$ has the real eigenvalues $p_j^2 = |p^B_j|^2 = 4/a_s^2\, \sin^2(p_j a_s/2)$. The magnitude of a lattice momentum will be denoted by $p = |\mbf p^B|= |\mbf p^F| \equiv \sqrt{\sum_j p_j^2}$ in the following. This allows us to define a basis of vectors $\mbf v^{(\lambda)}(\mbf p)$ that will be referred to as polarization vectors $\mbf v^{(\lambda)}(\mbf p)$ with polarization $\lambda$. The longitudinal polarization $\lambda = 3$ is $\mbf v^{(3)} (\mbf p)= \mbf p^F / p$, while the remaining vectors with $\lambda = 1,2$ are orthonormal to it and to each other and are commonly referred to as transversely polarized. One can now define transverse and longitudinal projection operators\footnote{Strictly speaking, $2 P^T$ is a projection operator. The additional factor $\nicefrac{1}{2}$ in the definition of $P^T$ corresponds to an average over transverse polarizations.} as
\begin{align}
 \label{eq_proj_long_trans}
 P^T_{ij}(\mbf p) &= \frac{1}{2}\sum_{\lambda = 1,2} v_i^{(\lambda)}(\mbf p) \left(v_j^{(\lambda)}(\mbf p)\right)^* \nonumber \\
 P^L_{ij}(\mbf p) &= v_i^{(3)}(\mbf p) \left(v_j^{(3)}(\mbf p)\right)^*
\end{align}
while one has $2 P^T_{ij}(\mbf p) + P^L_{ij}(\mbf p) = \delta_{ij}$. 

The gauge and electric fields are initialized in momentum space with only transverse polarizations as
\begin{align}
\label{eq_fields_init}
 A_j^a(t=0,\mbf p) &= \sqrt{\frac{f(t=0, p)}{p}} \sum_{\lambda = 1,2} c^{(\lambda)}_a(\mbf p)\, v_j^{(\lambda)}(\mbf p) \nonumber \\
 E^j_a(t=0,\mbf p) &= \sqrt{p\,f(t=0, p)} \sum_{\lambda = 1,2} \tilde{c}^{(\lambda)}_a(\mbf p)\, v_j^{(\lambda)}(\mbf p),
\end{align}
with complex Gaussian random numbers that satisfy
\begin{align}
\label{eq_random_nums}
 \left\langle \left(c^{(\lambda)}_{a}(\mbf p)\right)^* c^{(\lambda')}_{a'}(\mbf p') \right\rangle_{\cl} = V \delta_{\mbf p, \mbf p'} \delta_{a,a'} \delta_{\lambda,\lambda'},
\end{align}
and similarly for $\tilde{c}$, while $\langle \tilde{c}^*  c \rangle_\cl = 0$, where $\langle \cdot \rangle_{\cl}$ denotes a classical average over the distribution of these random numbers. 

We choose an isotropic initial single-particle distribution 
\begin{align}
\label{eq_init_cond}
 f(t=0, p) = \frac{n_0}{g^2}\, \frac{\pInit}{p}\, e^{-\frac{p^2}{2\pInit^2}}.
\end{align}
The fields constructed in this way are not guaranteed to satisfy the Gauss law~\nr{eq:gauss}, which thus has to be separately imposed by projecting back to the constraint surface using the algorithm in~\cite{Moore:1996qs}. The form (\ref{eq_init_cond}) of the initial conditions, with the initial amplitude $n_0$ and momentum scale $\pInit$, is expected to be close to the attractor distribution encountered in highly occupied plasmas \cite{York:2014wja}. Moreover, for weak coupling $g \ll 1$ this initial condition guarantees high occupation numbers and we always compute $g^2 f$, whose evolution is independent of the coupling constant in classical-statistical simulations since $g$ drops out of the classical equations of motion for these initial conditions. For convenience, we also define a characteristic energy scale
\begin{align}
 \Q = \sqrt[4]{5\,n_0}\; \pInit \propto \sqrt[4]{g^2 \epsilon},
\end{align}
where $\epsilon$ is the energy density. Unless stated otherwise, all dimensionful quantities will be expressed in terms of $\Q$.

\begin{figure}[t]
	\centering
	\includegraphics[scale=0.7]{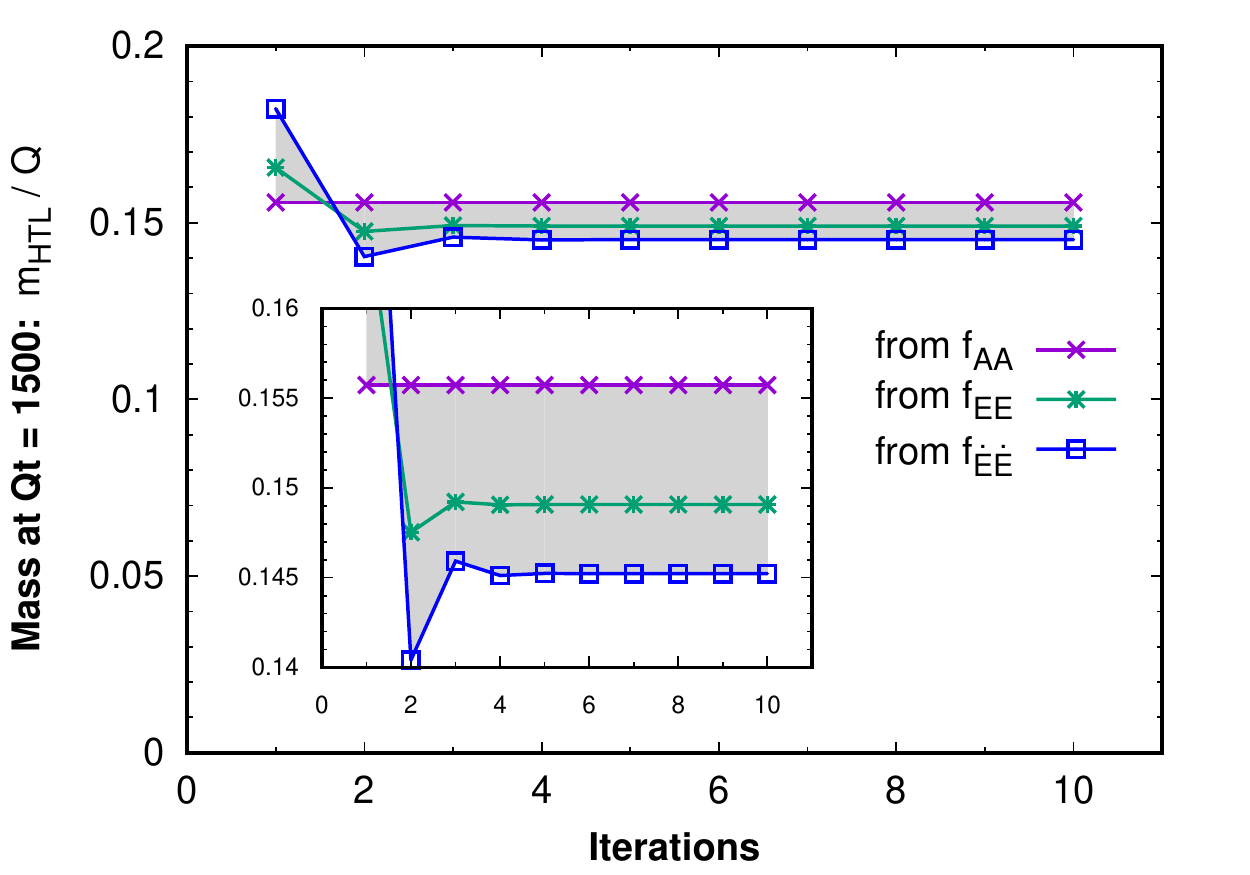}
	\caption{The mass (\ref{eq_mass_formula}) at $\Q t = 1500$ and $n_0 = 0.2$ for different definitions of the distribution function in \eqref{eq_distr_def}. In the inset, the values for the mass can be better read off.}
	\label{fig_msqr_fp_defs}
\end{figure}

At output times $t > 0$, we define the distribution function with transversely polarized fields as
\begin{align}
\label{eq_distr_def}
 f_{\mrm{EE}}(t,p) &= \frac{1}{d_A V}\,P_{ij}^T(\mbf p) \frac{\left\langle E^j_{a}(\mbf p) \left(E^i_{a}(\mbf p)\right)^* \right\rangle_\cl}{\sqrt{m_\HTL^2 + p^2}} \nonumber \\
 f_{\mrm{AA}}(t,p) &= \frac{\sqrt{m_\HTL^2 + p^2}}{d_A V}\, P^{ij}_T(\mbf p)\left\langle A_j^{a}(\mbf p) \left(A_i^{a}(\mbf p)\right)^* \right\rangle_\cl \nonumber \\
 f_{\mrm{\dot{E}\dot{E}}}(t,p) &= \frac{1}{d_A V}\,P_{ij}^T(\mbf p) \frac{\left\langle \partial_t E^j_{a}(\mbf p) \left(\partial_t E^i_{a}(\mbf p)\right)^* \right\rangle_\cl}{\left( m_\HTL^2 + p^2 \right)^{3/2}}
\end{align}
and employ $f(t,p) = f_{\mrm{EE}}(t,p)$ as our standard definition of the distribution function. In all definitions the distribution function is averaged over transverse polarizations and over adjoint gauge components, where $d_A = N_c^2 - 1$ is the dimension of the adjoint representation. Since the considered systems are isotropic, we additionally average over momentum modes with the same magnitude $p$ when computing observables, which improves the statistical accuracy of our results. The mass entering these expressions is the (asymptotic) mass $m$, computed iteratively as 
\begin{align}
\label{eq_mass_formula}
 m_\HTL^2 = 2 N_c \int \frac{\mrm d^3 p}{(2\pi)^3}\,\frac{g^2 f(t,p)}{\sqrt{m_\HTL^2 + p^2}}\,,
\end{align}
where we refer to it as $m_\HTL$ to indicate its connection to the HTL formalism. While in the corresponding HTL expression at LO the mass does not occur on the right hand side, Eq.~\eqref{eq_mass_formula} takes some NLO corrections into account. 

The different distribution functions defined in \eqref{eq_distr_def} can lead to slightly different results in HTL computations. This is demonstrated at the example of the mass $m_\HTL$ that is shown in Fig.~\ref{fig_msqr_fp_defs} as a function of the iteration step for different definitions of the distribution function. In the first iteration, we use $m_\HTL = 0$, which corresponds to the LO formula for the mass. Since $\sqrt{m_\HTL^2 + p^2}$ drops out of the calculation of the mass when $f_{\mrm{AA}}$ is used, the corresponding value does not change with the iterations, while $m_\HTL$ computed with other definitions jumps to a lower value. One observes that the value for $m_\HTL$ ranges between $0.145\, \Q$ and $0.156\, \Q$, with $f_{\mrm{EE}}$ leading to a value in-between
\begin{align}
 \label{eq_mHTL_value}
 m_\HTL(\Q t = 1500) = 0.149 \, \Q.
\end{align}
We use $m_\HTL$ computed with the distribution function $f_{\mrm{EE}}$ for all plots where $m_\HTL$ is used in \se\ref{sec:results} because it typically provides values for the mass located between the ones from the other definitions. However, we emphasize that this value is not precise and other definitions provide values that differ by typically $5\, \%$. Similarly, we use the spread in other HTL computed observables to estimate its error.

\subsection{Spectral and statistical correlation functions}
\label{sec_corr_fcts}

We are mainly interested in properties of two-point correlation functions, especially at unequal time, which encode information about the quasiparticle character, the excitation spectrum and the distribution of quasiparticles. 

For this purpose, one can define the statistical correlation function and its double time derivative, which are symmetric and real-valued functions, as anticommutators of two field operators \cite{Berges:2004yj}
\begin{align}
 \label{eq_stat_fct}
 F_{jk}^{ab} (x, x') &= \frac{1}{2}\left\langle \left\{ \hat{A}_j^a(x), \hat{A}_k^b(x') \right\} \right\rangle \nonumber \\
 \ddot{F}^{jk}_{ab} (x, x') &= \frac{1}{2}\left\langle \left\{ \hat{E}^j_a(x), \hat{E}^k_b(x') \right\} \right\rangle,
\end{align}
with $x = (t,\mbf x)$ and where we used that $\partial_t \hat{A}_i^b(x) = \hat{E}^i_b(x)$ (in the continuum). Since we consider spatially homogeneous systems, $F$ does not depend on the central spatial coordinates $(\mbf x + \mbf x')/2$ but only on the relative coordinates $\Delta \mbf x = \mbf x - \mbf x'$ and one can perform a spatial Fourier transform with respect to $\Delta \mbf x$, arriving at $F(t,t',\mbf p)$ and $\ddot{F}(t,t',\mbf p)$. Further employing the system's isotropy, one arrives at $F(t,t',p)$, $\ddot{F}(t,t',p)$. 

In classical-statistical simulations, the anticommutator of Heisenberg fields reduces to a product of classical fields $\nicefrac{1}{2}\, \langle \{ \hat{A}, \hat{A'} \} \rangle \mapsto \langle A\, A' \rangle_\cl$. Hence, the definitions (\ref{eq_stat_fct}) become
\begin{align}
 \label{eq_stat_fct_class}
 F (t,t', p) &= \frac{1}{d_A V}\,P^{kj}(\mbf p)\left\langle A_j^b(t, \mbf p)\,\left(A_k^b(t', \mbf p)\right)^* \right\rangle_\cl  \nonumber \\
 \ddot{F} (t,t', p) &= \frac{1}{d_A V}\,P_{kj}(\mbf p)\left\langle E^j_b(t, \mbf p)\,\left(E^k_b(t', \mbf p)\right)^* \right\rangle_\cl,
\end{align}
where we employed the operators $P^{T/L}_{jk}$ from \eqref{eq_proj_long_trans} to average over transverse or longitudinal polarizations but to simplify notation, we omitted the subscript $T$ or $L$. Both $F$ and $\ddot{F}$ also include averaging over the $d_A = N_c^2 - 1$ adjoint gauge components. From \eq\eqref{eq_stat_fct_class} it becomes apparent that the statistical correlation function is closely related to the distribution function. Indeed, our definitions of the latter in \eq(\ref{eq_distr_def}) can be expressed as $f_{\mrm{EE}}(t,p) = \ddot{F}_T(t,t, p) / \sqrt{m_\HTL^2 + p^2}$ and similarly for the others. 

Another important correlation function is the spectral function (and its time derivative) that can be defined as the commutator of two field operators
\begin{align}
 \label{eq_spectral_fct}
 \rho_{jk}^{ab} (x, x') &= i \left\langle \left[ \hat{A}_j^a(x), \hat{A}_k^b(x') \right] \right\rangle  \nonumber \\
 \dot{\rho}_{jk}^{ab} (x, x') &= i \left\langle \left[ \hat{E}^j_a(x), \hat{A}_k^b(x') \right] \right\rangle. 
\end{align}
Following the same steps as provided below \eq(\ref{eq_stat_fct}), one arrives at 
\begin{align}
 \label{eq_spectral_fct_class}
 \rho (t,t', p) &= \frac{1}{d_A}\,P^{kj}(\mbf p)\,\rho_{jk}^{b b}(t,t',\mbf p) \nonumber \\
 \dot{\rho} (t,t', p) &= \frac{1}{d_A}\,P^{kj}(\mbf p)\, \dot{\rho}_{jk}^{b b}(t,t',\mbf p)\,.
\end{align}
The spectral function is anti-symmetric and real valued. The equal-time relations of the transverse component are fully determined by the canonical commutation relations
\begin{align}
 \label{eq_rho_init_cond}
 \lim_{t \rightarrow t'} \rho_T(t, t', p) &= 0 \nonumber \\
 \lim_{t \rightarrow t'} \dot{\rho}_T(t, t', p) &= 1.
\end{align}
While the first relation is also valid for the longitudinal component $\lim_{t \rightarrow t'} \rho_L(t, t', p) = 0$, the equivalent of the second relation will be discussed in \se\ref{sec_HTL_theory}. These transverse and longitudinal equal-time relations for the spectral function fully determine the initial conditions for its evolution. Note that this is different from the statistical correlation function whose initial conditions can be chosen by specifying an initial distribution $f(t=0,p)$. 

Since in a classical-statistical framework commutators are mapped to Poisson-brackets (or Dirac-brackets in our case), a direct computation of the spectral function as in the case of the statistical correlator is more involved. Instead, we will use linear response theory for its computation, which will be discussed in \se\ref{sec_linresp}. There it is described how one can compute the retarded propagator $G_R(t,t',p)$, which is closely related to the spectral function via
\begin{align}
 \label{eq_GR_rho_relation}
 G_{R}(t,t',p) = \theta(t - t')\, \rho(t,t',p). 
\end{align}
Thus, having computed $G_R$ and exploiting the fact that for positive time differences it coincides with $\rho$, we simultaneously get the spectral function. 

To discuss quasiparticle excitations, it is useful to transform the correlation functions to frequency space. This is done by first  transforming the time variables to $\tcent = (t + t')/2$ and $\Delta t = t - t'$ and Fourier transforming the correlation functions with respect to $\Delta t$ while keeping $\tcent$ fixed. Ideally, we would compute 
\begin{align}
 \label{eq_F_rho_w_def}
 F(\tcent,\omega, p) =\, & 2\int_{0}^{\infty} \ud \Delta t \, \cos( \omega\, \Delta t)\, F(\tcent +\Delta t/2, \tcent-\Delta t/2, p) \nonumber \\
 \rho(\tcent,\omega, p) =\, & 2\int_{0}^{\infty} \ud \Delta t \,\sin( \omega\, \Delta t)\, \rho(\tcent +\Delta t/2, \tcent-\Delta t/2, p),
\end{align}
but in the interest of practicality we approximate this by 
\begin{align}
 \label{eq_F_rho_w_comp}
 F(\tcent,\omega, p) \approx\, & 2\int_{0}^{\dtmax} \ud \Delta t \, \cos( \omega\, \Delta t)\, F(\tcent +\Delta t, \tcent, p) \nonumber \\
 \rho(\tcent,\omega, p) \approx\, & 2\int_{0}^{\dtmax} \ud \Delta t \,\sin( \omega\, \Delta t)\, \rho(\tcent +\Delta t, \tcent, p),
\end{align}
and always replace $\tcent$ by $t$. This approximation is justified when taking a sufficiently large time $t$ such that $\dtmax \ll t$ and $\tcent \approx t$. Moreover, values at small $\Delta t$ typically give the dominant contributions to the integral because of the damping of oscillations in the correlation functions. Note that we have defined a real-valued spectral function in frequency space in Eqs.~\eqref{eq_F_rho_w_def} and \eqref{eq_F_rho_w_comp} by dropping a factor of $i$. 

Because of the separation of (time) scales $\tcent^{-1} \ll \omega$, where $\omega$ is a typical frequency we are interested in, the correlation functions change much faster as functions of relative time $\Delta t$ than of central time $\tcent$. Thus, $\omega \rho$  is approximately $\dot{\rho}$ and similarly, the statistical correlation function $\omega^2 F$ becomes $\ddot{F}$.\footnote{Indeed, we checked that $\omega \rho$ and $\dot{\rho}$, and similarly $\omega^2 F$ and $\ddot{F}$, lie on top of each other in frequency space to high accuracy, which will also be shown explicitly in \fig\ref{fig_lorentz}.}

\subsection{Predictions from HTL theory}
\label{sec_HTL_theory}

Having introduced the relevant correlation functions, we provide a short summary of expressions that are derived within the HTL formalism at LO \cite{Braaten:1989mz,lebellac_1996}. Although developed primarily for thermal equilibrium, the HTL formalism can also be applied to systems out of equilibrium \cite{Mrowczynski:2000ed,Arnold:2002zm}. 

An important quantity is the polarization tensor whose transverse and longitudinal components are functions of the ratio $x = \omega/p$. For isotropic non-equilibrium systems with a scale separation between $m$ and the hard scale $\Lambda$, the polarization tensors are given by
\begin{align}
 \Pi_T(x) &= m^2\, x\left( x + (1-x^2)Q_0(x) \right) \nonumber \\
 \Pi_L(x) &= -2m^2\, \left(1 - x\,Q_0(x)\right), 
\end{align}
with the Legendre function of the second kind
\begin{align}
 \label{eq_Legendre_fct}
 Q_0(x) = \frac{1}{2}\ln \frac{x+1}{x-1} = \frac{1}{2}\ln \left|\frac{x+1}{x-1}\right| -\frac{i\pi}{2} \theta (1-x^2). 
\end{align}
At strict leading order, the mass $m$ can be computed by the LO version of the HTL expression \eqref{eq_mass_formula}. While the polarization tensor is gauge invariant at LO, the exact form of the retarded propagator $G_R^{\HTL}$ depends on the chosen gauge. For the temporal gauge $A_0 = 0$ as here employed, its transverse and longitudinal components read
\begin{align}
 \label{eq_GR_HTL}
 G_T^{\HTL}(\omega, p) &= \frac{-1}{\omega^2 - p^2 - \Pi_T(\omega/p)} \nonumber \\
 G_L^{\HTL}(\omega, p) &= \frac{p^2}{\omega^2}\, \frac{-1}{p^2 - \Pi_L(\omega/p)}\,.
\end{align}
The components of the spectral function are obtained by 
\begin{align}
 \label{eq_rhoHTL_GT_rel}
 \rho^{\HTL} (\omega, p) = 2\, \mrm{Im}\,G^{\HTL}(\omega, p),
\end{align}
which is consistent with our formerly stated relations in \eqref{eq_GR_rho_relation} and \eqref{eq_F_rho_w_comp}. We note that in the literature the longitudinally polarized spectral function is usually defined as
\begin{align}
 \tilde{\rho}_L^\HTL (\omega, p) = \frac{\omega^2}{p^2}\,  \rho_L^{\HTL} (\omega, p)
\end{align}
to compensate for the gauge dependent prefactor $p^2 / \omega^2$ in \eqref{eq_GR_HTL}. However, we will use $\rho_L^{\HTL}$ instead of $\tilde{\rho}_L^{\HTL}$ in comparisons with our data since $\rho_L$ is what we obtain in both time and frequency domains numerically. 

Quasiparticles emerge as poles of the retarded propagator. Therefore, by finding the roots of the denominator in \eqref{eq_GR_HTL} numerically, one obtains the transverse and longitudinal dispersion relations $\omega_{T/L}^{\HTL}(p)$. Their low- and high-momentum expansions read
\begin{align}
 \label{eq_wHTL_Taylor}
 \omega_{T}^{\HTL} &\overset{p\, \ll\, m}{\simeq} \sqrt{(\wplas^\HTL)^2 + 1.2\,p^2} \nonumber \\
 \omega_{L}^{\HTL} &\overset{p\, \ll\, m}{\simeq} \sqrt{(\wplas^\HTL)^2 + 0.6\,p^2} \\
 \omega_{T}^{\HTL} &\overset{p\, \gg\, m}{\simeq} \sqrt{m^2 + p^2} \nonumber \\
 \omega_{L}^{\HTL} &\overset{p\, \gg\, m}{\simeq} p \left( 1 + 2\,\exp\left( -\frac{m^2 + p^2}{m^2} \right) \right). 
\end{align}
Here the plasmon frequency in the HTL framework is given by 
\begin{align}
 \label{eq_wplas_m_rel}
 \wplas^\HTL = \sqrt{2/3}\;m\,,
\end{align}
and is approached for both dispersion relations in the limit of low momenta. At high momenta, the transverse dispersion relation corresponds to a relativistic dispersion with asymptotic mass $m$ while for longitudinal momenta, one essentially has an ultrarelativistic dispersion $\omega_{L}^{\HTL} \approx p$. 

The quasipartice peak enters the spectral function as a Delta function $\delta(\omega - \omega_{T/L}^{\HTL})$ with a prefactor. Due to the imaginary part of the Legendre function \eqref{eq_Legendre_fct}, the polarization tensor obtains an imaginary part for low frequencies $\omega^2 \leq p^2$, which also enters the spectral function and is referred to as the Landau cut. Hence, we can write the spectral function as a sum of the Landau cut region $\beta_{T,L}(\omega, p)$ and the quasiparticle peak
\begin{align}
 \label{eq_rhoT_HTL}
 \rho_{T}^{\HTL} (\omega, p) &= \beta_{T}(\omega, p)  + \, \mrm{q.p. peak} \nonumber \\
 \tilde{\rho}_{L}^{\HTL} (\omega, p) &= \beta_{L}(\omega, p)  + \, \mrm{q.p. peak}. 
\end{align}
The Landau cut region with $x = \omega / p$ is given by 
\begin{align}
 &\beta_T(\omega, p) = \pi m^2  x (1-x^2)\, \theta (1-x^2) \nonumber \\
 &\times \Bigg[\left( p^2 (x^2 - 1) - m^2 \left( x^2 + \frac{1}{2}\, x (1-x^2) \ln \left|\frac{1+x}{1-x}\right| \right) \right)^2 \nonumber \\
 &\qquad \qquad + \left( \frac{\pi}{2}\, m^2 x (1-x^2) \right)^2 \Bigg]^{-1},
\end{align}
for the transverse case and by 
\begin{align}
 \label{eq_Landau_cut_long}
 &\beta_L(\omega, p) = 2\pi m^2  x \, \theta (1-x^2) \nonumber \\
 &\times \Bigg[\left( p^2 + 2m^2 \left( 1 - \frac{x}{2}\, \ln \left|\frac{1+x}{1-x}\right| \right) \right)^2 + \left( \pi\, m^2 x \right)^2 \Bigg]^{-1},
\end{align}
for the longitudinal polarization. The spectral function satisfies the sum rules
\begin{align}
 \label{eq_rho_sum_rules}
 \dot{\rho}_T^\HTL (\Delta t = 0,p) &= 2\int_{0}^\infty \frac{\ud \omega}{2\pi}\,\dot{\rho}_T^\HTL (\omega,p) = 1 \nonumber \\
 \dot{\rho}_L^\HTL (\Delta t = 0,p) &= 2\int_0^{\infty} \frac{\ud \omega}{2\pi}\, \dot{\rho}_L^\HTL (\omega, p) = \frac{2m^2}{2m^2 + p^2}.
\end{align}
The transverse sum rule is equivalent to \eqref{eq_rho_init_cond} while the longitudinal sum rule results from $\int_{-\infty}^\infty \ud \omega/(2\pi)\, \tilde{\rho}_L^\HTL (\omega, p)/\omega = 2m^2 / (p^2(2m^2 + p^2))$.

The HTL framework also predicts that the statistical correlation function $F$ is not independent of the spectral function $\rho$. Instead,  they are connected at soft momenta and frequencies $\omega, p \ll \Lambda$ via 
\begin{align}
 \label{eq_fluct_diss_rel_HTL}
 F^\HTL(\tcent, \omega, p) = \frac{T_*(\tcent)}{\omega}\, \rho^{\HTL} (\tcent, \omega, p)\,,
\end{align}
with $T_*$ given by
\begin{align}
 T_*(t) = \mathcal{I}(t) / \mathcal{J}(t)
\end{align}
and the integrals 
\begin{align}
 \mathcal{I}(t) &= \frac{1}{2} \int \frac{\ud^3 p}{(2\pi)^3}\, f(t,p) \left(f(t,p) + 1 \right) \nonumber \\
 \mathcal{J}(t) &= \int \frac{\ud^3 p}{(2\pi)^3}\, \frac{f(t,p)}{\sqrt{m_\HTL^2 + p^2}}\,,
\end{align}
where we generalized $g^2\mathcal{J} = m_\HTL^2/(2N_c)$ to include the self-consistent resummation of the frequency denominator $f/p \mapsto f / \sqrt{m_\HTL^2 + p^2}$, as we did for the mass. Moreover, for systems with large occupation numbers $f(t,\Lambda) \gg 1$ one can make the approximation $f (f+1) \approx f^2$. On the other hand, if $f$ is given by the thermal Bose-Einstein distribution $(e^{\omega/T} - 1)^{-1}$ with $\omega \approx p$, one obtains $T = T_*$ and Eq.~\eqref{eq_fluct_diss_rel_HTL} becomes the fluctuation-dissipation relation in thermal equilibrium for $\omega \ll \Lambda$. The relation \eqref{eq_fluct_diss_rel_HTL} can be turned to an equal-time relation by multiplying the equation by $\omega^2$ and integrating over the frequency, which leads to
\begin{align}
 \label{eq_fluct_diss_equTime_HTL}
 \ddot{F}^\HTL(\tcent, \Delta t = 0, p) = T_*(\tcent)\; \dot{\rho}^{\HTL} (\tcent, \Delta t = 0, p)\,. 
\end{align}
If the distribution function follows a $1/p$ power law at momenta $p \ll \Lambda$, then $T_*$ can be understood as the effective temperature of low momenta where $f(t,p) \approx T_*/p$. 

While in a thermal system $m$ and $T$ are constants in time, in a nonequilibrium system the distribution function depends on time $\tcent$, which for instance implies a slow time dependence of the mass \eqref{eq_mass_formula}. Important assumptions for this are an existing scale separation between the hard scale $\Lambda$ which dominates the energy density and the asymptotic mass $m$ as well as that relevant loop diagrams are dominated by modes of the order of $\Lambda$. Although in general this may pose constraints for the form of the distribution function, for the case of a highly occupied non-Abelian plasma close to its self-similar scaling solution that will be studied in \se\ref{sec:results} these conditions are expected to be satisfied up to higher order effects.

\subsection{Linear response theory}
\label{sec_linresp}

To compute the retarded propagator $G_R$ numerically, we study the response of the non-Abelian plasma to a small perturbation with a source $j^k_b(x) = j^k_b(t, \mbf x)$. Then the plasma field can be split into two parts 
\begin{align}
 \hat{A}_k^b \rightarrow \hat{A}_k^b + \hat{a}_k^b, \quad \hat{E}^k_b \rightarrow \hat{E}^k_b + \hat{e}^k_b,
\end{align}
which are written as field operators in the Heisenberg picture. If no source is applied, the response is zero $\langle \hat{a}\rangle = \langle \hat{e}\rangle = 0$. Otherwise, it is given by\footnote{Note that we use a phase convention where $G_R(x,x')$ is real-valued.} \cite{lebellac_1996}
\begin{align}
\label{eq_lin_resp_basic}
 \langle \hat{a}_i^b(x)\rangle = \int \mrm{d}^4 x' G_{R,ik}^{~~bc}(x, x')\, j^k_c(x'),
\end{align}
with the retarded propagator
\begin{align}
 G_{R,ik}^{~~bc}(x, x') = i\theta(t-t')\, \left\langle \left[ \hat{A}_i^b(x), \hat{A}_k^c(x') \right] \right\rangle. 
\end{align}
Since we consider spatially homogeneous systems, $G_R$ does not depend on the central spatial coordinates $(\mbf x + \mbf x')/2$ but only on the relative coordinates $\Delta \mbf x = \mbf x - \mbf x'$. In Fourier space, Eq.~(\ref{eq_lin_resp_basic}) then reads
\begin{align}
 \langle \hat{a}_i^b(t, \mbf p)\rangle = \int \ud t'\, G_{R,ik}^{~~bc}(t, t', \mbf p)\, j^k_c(t',\mbf p).
\end{align}
Using for the source an instant perturbation of a mode $\mbf p$ at time $\tpert$
\begin{align}
\label{eq_source_init_cont}
 j^k_c(t',\mbf p) = j^k_{0,c}(\mbf p)\, \delta\left( t' - \tpert \right),
\end{align}
one arrives at
\begin{align}
 \langle \hat{a}_i^b(t, \mbf p)\rangle= G_{R,ik}^{~~bc}(t, \tpert, \mbf p)\, j^k_{0,c}(\mbf p).
\end{align}
From this, we would like to compute the retarded propagator $G_R$. Because of the summation over indices, we cannot simply divide this expression by the source term. Moreover, since we consider a linear response to a perturbation, we can set multiple momentum modes simultaneously for the source $j_{0}$. To deduce the retarded propagator, we choose the perturbation to satisfy~\footnote{We are grateful to A.~Pi\~{n}eiro Orioli for sharing with us this method in a private communication. This method has been developed for non-relativistic scalar field theories in Ref.~\cite{Orioli:2018xxx}.}
\begin{align}
\label{eq_source_relations}
\left\langle j^k_{0,b}(\mbf p)\, \left(j^{k'}_{0,b'}(\mbf p')\right)^*\right\rangle_\uj = \delta_{b,b'}\; V \delta_{\mbf p, \mbf p'}\; d_\lambda P^{k\, k'}_{T/L}(\mbf p),
\end{align}
where the number of polarizations is $d_\lambda = 2$ for the transverse and $d_\lambda = 1$ for the longitudinal case and where $\left\langle \cdot \right\rangle_\uj$ is a classical average over a set of sources. Indeed, similar to the initial conditions in our simulations in \eq(\ref{eq_fields_init}), the relations (\ref{eq_source_relations}) can be achieved by choosing 
\begin{align}
 \label{eq_j_init}
 j^k_{0,b}(\mbf p) = \sum_{\lambda} c^{(\lambda)}_b(\mbf p)\, v_k^{(\lambda)}(\mbf p),
\end{align}
with a random phase $c^{(\lambda)}_b(\mbf p)$ satisfying (\ref{eq_random_nums}) with the replacement $\langle . \rangle_\cl \rightarrow \langle . \rangle_\uj$. The summation over polarizations $\lambda$ in the initialization of the source is chosen to involve only transverse modes ($\lambda = 1,2$) when discussing the transversely projected spectral and retarded correlation functions and only the longitudinal polarization ($\lambda = 3$) in the longitudinal case. As usual, we will occasionally omit the subscripts $T, L$ to simplify notation. 

The retarded propagator, averaged over adjoint components and polarizations, can then be computed as
\begin{align}
 G_{R}(t, \tpert, \mbf p) &= \frac{1}{d_A d_\lambda V}\, \left\langle \langle \hat{a}_i^b(t, \mbf p)\rangle \left(j^i_{0, b}(\mbf p)\right)^* \right\rangle_\uj \nonumber \\
 &= \frac{1}{d_A}\,P^{ki}(\mbf p)\, G_{R,ik}^{~~bb}(t, \tpert, \mbf p).
\end{align}
Similarly, we define the time derivative of the retarded propagator as
\begin{align}
 \dot{G}_{R}(t, \tpert, \mbf p) = \frac{1}{d_A d_\lambda V} \left\langle \langle \hat{e}^k_b(t, \mbf p)\rangle \left(j^k_{0, b}(\mbf p)\right)^* \right\rangle_\uj,
\end{align}
such that one has $\dot{G}_{R}(t,t',p) = \theta(t - t')\, \dot{\rho}(t,t',p)$. Therefore, and due to the relation \eqref{eq_GR_rho_relation}, the retarded propagator is fixed at $t \rightarrow \tpert$ in the same way as the spectral function, which serves as the initial condition for its evolution.

\subsection{Linearized fluctuations}
\label{sec_linflucts}

To compute the linear response in the discretized classical-statistical framework, the gauge and chromoelectric fields are split according to 
\begin{align}
 A_k^b \rightarrow A_k^b + a_k^b, \quad E^k_b \rightarrow E^k_b + e^k_b,
\end{align}
where $A_k^b$ and $E^k_b$ will from now on be referred to as background fields and $a_k^b$ and $e^k_b$ are the linearized fluctuation fields. They transform under a gauge transformation $V(x)$ in the same way as the electric background field, which is $E^k_a(\mbf x) \rightarrow V^{ab}(\mbf x) E^k_b(\mbf x)$ in the adjoint representation. These fluctuations are the expectation values of the fluctuation field operators of \se\ref{sec_linresp}, where in the classical approximation the  expectation value is calculated as an average of the classical distribution
\begin{align}
 \langle \hat{a}_k^b \rangle \rightarrow \langle a_k^b \rangle_\cl, \quad \langle \hat{e}^k_b \rangle \rightarrow \langle e^k_b \rangle_\cl,
\end{align}
and the formulas for $G_{R}$ and $\dot{G}_{R}$ have to be changed accordingly. 

The equations of motion for the fluctuations are derived in Ref.~\cite{Kurkela:2016mhu} by linearizing the lattice equations of motion (\ref{eq_EOM_BG}) and additionally demanding that the Gauss law condition (\ref{eq:gauss}) is satisfied in the linearized framework. For an $\mrm{SU}(2)$ theory, the $a$-field update is given by
\begin{align}
 \label{eq_EOM_linflucts}
 a_j^b(t + \ud t / 2, \mbf x) =\; & a_j^{\parallel, b} + \ud t\, e^j_{\parallel, b} + U_{0j}^{bc}\, a_j^{\bot, c} \nonumber \\
 + \frac{\epsilon^{bcd}}{a_s\,(gE^j)^2}&\; gE^j_c \left( U_{0j}^{da}\,e^j_{\bot, a} - e^j_{\bot, d} \right),
\end{align}
where we defined the matrix $U_{0j}(t, \mbf x) = \exp( i\,\ud t\, a_s gE^j_a(t,\mbf x)\Gamma^a)$. All electric fields on the right hand side are evaluated at $(t, \mbf x)$ while the linearized gauge field appears there at $(t - \ud  t/2, \mbf x)$. Moreover, we split the linearized fields in parts parallel and transverse to the electric background field in color space, $e^j_b = e^j_{\parallel, b} + e^j_{\bot, b}$, with $\sum_b e^j_{\bot, b} E^j_{b} = 0$ for all $j$ and $(E^j)^2 = \sum_b E^j_{b} E^j_{b}$. Writing $a_j(t,\mbf x) = a_j^b(t,\mbf x)\Gamma^b$, the $e$-field update reads
\begin{align}
  e^j(t\, +&\, \ud t ,\mbf x) = e^j(\mbf x) + \ud t\, j^j(\mbf x) \nonumber \\
 - \frac{\ud t}{a_s^2} \sum_{k \neq j} i\Big[& \Big(a_j(\mbf x) + a_k(\mbf x + \jj \rightarrow \mbf x) \Big)U_{jk}(\mbf x) \nonumber \\
 -\,&U_{jk}(\mbf x) \Big(a_j(\mbf x + \kk \rightarrow \mbf x) + a_k(\mbf x) \Big) \nonumber \\
 + \Big(a_j(\mbf x)\, - &\,a_k(\mbf x + \jj -\kk \rightarrow \mbf x + \jj \rightarrow \mbf x) \Big)U_{j(-k)}(\mbf x) \nonumber \\
 -\, U_{j(-k)}(\mbf x)& \Big(a_j(\mbf x -\kk \rightarrow \mbf x) - a_k(\mbf x -\kk \rightarrow \mbf x) \Big) \Big]_{\ah},
\end{align}
where electric fields and the source $j$ on the right hand side are at time $t$ while the gauge and link fields are taken at $t + \ud t / 2$. For parallel transported fields we used the notation $a_k(\mbf x + \jj \rightarrow \mbf x) = U_j(\mbf x) a_k(\mbf x + \jj) U_j^\dagger(\mbf x)$. 

We discuss now the initial conditions of the linearized fields. Before the time of the perturbation $\tpert$, the linearized fields $a$ and $e$ are zero. Using a source term according to (\ref{eq_source_init_cont})
\begin{align}
 j^k_b(t,\mbf p) = j^k_{0,b}(\mbf p)\, \frac{\delta_{t,\tpert- \ud t}}{\ud t},
\end{align}
the electric field at time $\tpert$ becomes
\begin{align}
 \label{eq_e_eq_j}
 e^k_b(\tpert,\mbf p) =  j^k_{0,b}(\mbf p), 
\end{align}
while the gauge field stays zero $a_k^b(\tpert,\mbf p) = 0$. Therefore, initializing just the linearized $e$ field is equivalent to perturbing the system with a source $j$ at time $\tpert$. Moreover, it can be easily checked that these initial conditions also satisfy the initial relations for the transverse retarded propagator and thus, also of the spectral function (\ref{eq_rho_init_cond}).\footnote{The initial condition for $\dot{G}_T$ is satisfied exactly before the Gauss law restoration algorithm is applied. After its application, however, this is not the case any more, especially for longitudinally polarized modes, and the initial conditions have to be set by hand by rescaling the amplitude accordingly for each momentum mode. This is possible for well separated excited momentum modes because the equations are linear for the fluctuations.}

To measure gauge-dependent observables such as momentum space correlations, one needs to fix the gauge.
The interpretation in terms of physical degrees of freedom is the clearest in a Coulomb gauge
\begin{align}
 \label{eq_Coulomb_gauge}
 \partial^B_j A_j = 0.
\end{align}
This implies that the gauge field is always transversely polarized while the electric field may have longitudinal contributions. 
For equal-time correlation functions the gauge is fixed always at the time of measurement, as for the distribution function $f(t,p)$. Here, however, we are measuring unequal-time correlators of the fields and the fluctuations. Fixing the gauge separately at each measurement time would mean that the fields at the different times would be taken from different gauge trajectories. As we show in Appendix~\ref{app:gaugefix}, this leads to gauge artefacts in the spectral function outside of the quasiparticle peak.  In order to avoid this effect, we only fix the Coulomb gauge condition at the time $\tpert$ when the fluctuation is introduced, but not at later times. Thus the system gradually shifts away from the gauge condition \eqref{eq_Coulomb_gauge}. However, the timescale of this deviation (which is of order $\tcent$) is expected to be long compared to the relevant time separations $\Delta t$ and the effect should therefore not be large. A more thorough investigation of the gauge dependence of our results is left for a future study. 
The gauge fixing is efficiently done with a Fourier-accelerated algorithm \cite{Davies:1987vs}.

Because of (\ref{eq_e_eq_j}), a sensible source $j$ has to also satisfy the Gauss law, whose violation in the linearized framework of Ref.~\cite{Kurkela:2016mhu} is defined by 
\begin{align}
 c^a(t, \mbf x ) =  &\hat{D}^{ab}_k\, e^k_b(t, \mbf x ) - \sum_k U_k^{\dagger,ab}(t, \mbf x - \hat{\mbf k})  \nonumber \\
 & \qquad \times f^{bcd} a_k^c(t, \mbf x - \hat{\mbf k})\, gE^k_d(t, \mbf x - \hat{\mbf k}) 
\end{align}
and is preserved by the linearized equations of motion. Therefore, to restore the Gauss law after the perturbation at $\tpert$, we employ the Gauss law restoration algorithm of Ref.~\cite{Moore:1996qs} for the Gauss violation $c^a$ only at time $\tpert$. The algorithm proceeds by iterating the transformation $e^k_a(\mbf x) \mapsto e^k_a(\mbf x) + \gamma \hat{D}^{F,ab}_k c^b(x)$, with the corresponding forward derivative $\hat{D}^{F,ab}_k$ and the same parameter $\gamma$ as in \cite{Moore:1996qs}. Our Gauss law violation is typically reduced to $10^{-9} - 10^{-12}$ in suitable units of $\Q$, which is close to machine precision. Our results are observed to be insensitive to the chosen Gauss law precision.

\subsection{Numerical setup}

In the following sections we show our numerical results. Our standard choice of parameters is $n_0 = 0.2$ and $\Q \tpert = 1500$, while our time extent for the Fourier transform is typically $\Q \dtmax = 200$, see also Sec.~\ref{sec_linresp}. If not stated otherwise, we employ a $256^3$ lattice with lattice spacing $\Q a_s = 0.7$, averaged over $5$ simulations. To show that our results are insensitive to changes of volume and lattice spacing, in some figures we compare to a smaller $192^3$ lattice with the finer spacing $\Q a_s = 0.47$, averaged over $10$ simulations. 
For the time step we use the ratio $\ud t / a_s = 0.05$, while we have checked that smaller ratios down to $\ud t / a_s = 0.01$ do not change the results. Unless stated otherwise, we initialize only transverse modes. 

Using the method described in Sec.~\ref{sec_linresp}, we initialize the source at multiple momenta $\mbf p$. To gain more statistics, we additionally bin our correlation functions linearly in momentum $p = |\mbf p^B|$, with the bin size being $1/8$ of the smallest momentum on the lattice $p_{\text{min}} = 2/a_s\, \sin(\pi/N_s)$. We also checked that smaller bin fractions do not change the results. 

We initialize all modes within a momentum bin, which amounts in initializing a thin spherical shell with radius $p$ in momentum space. Moreover, we do not initialize all momentum modes of a linearized fluctuation field. The reason for this is that, especially for low momentum modes, the evolution of $G_R$ becomes more noisy when neighboring momentum bins are also excited, since the initialized modes typically disperse and interfere with noise from dispersing neighboring modes if the latter were also excited. Achieving accurate results would require more statistics in that case. Therefore, we initialize single momentum bins separated by at least $0.15\,\Q$, thus reducing noise from neighboring points considerably. As noted above, our simulation results are averaged over $5 - 10$ realizations, where each realization has an independent background and source. The error bars correspond to the standard error of the mean. We will explicitly note where data without further averages is taken. 

Finally, to further reduce computational costs, we initialize multiple linearized fluctuations independently, exciting different modes, but with the same background field. In this way, we can extract the entire spectrum from a single simulation by initializing sufficiently many linearized fluctuation fields that most momentum bins become excited.

\section{Numerical results}
\label{sec:results}


\subsection{Self-similar attractor in non-Abelian plasmas}
\label{sec_selfsim_attr}

\begin{figure}[t]
	\centering
	\includegraphics[scale=0.7]{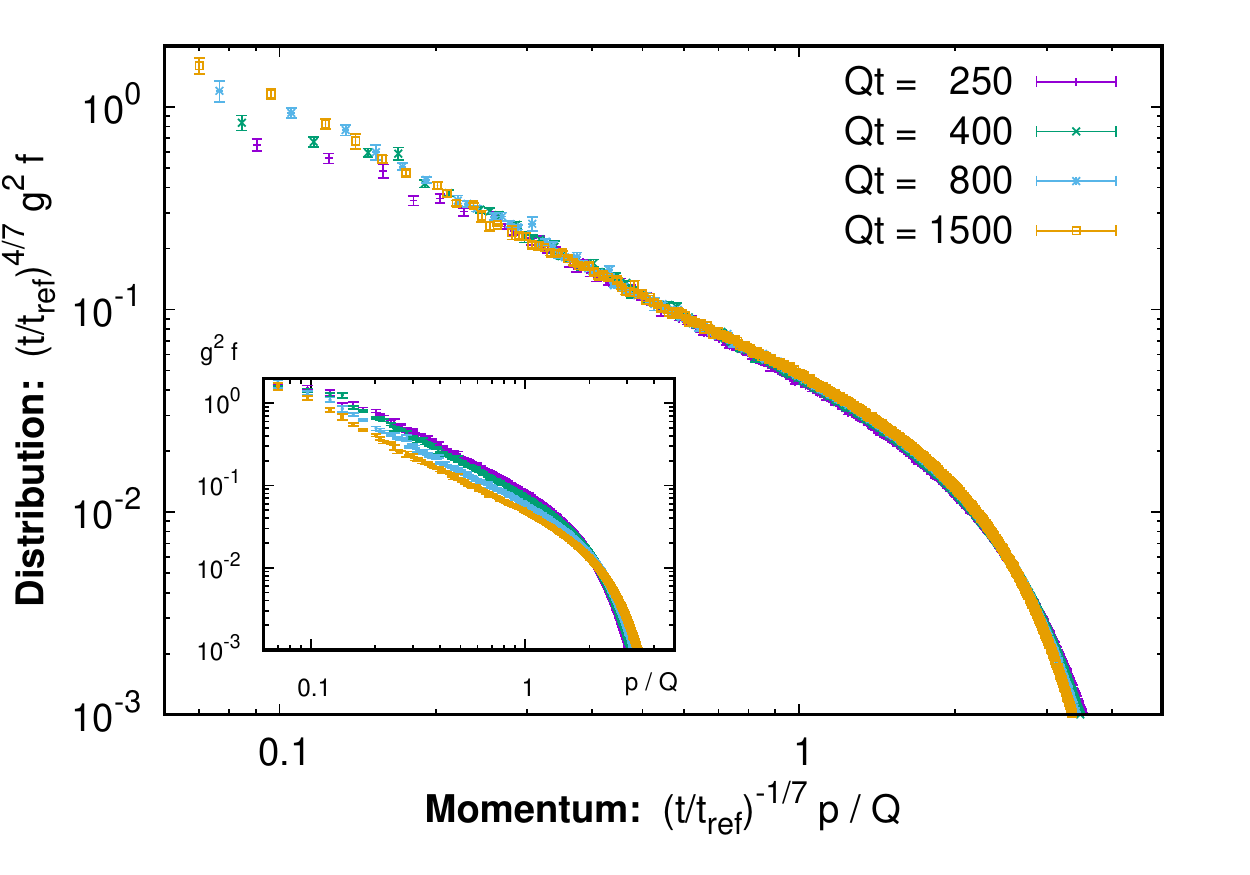}
	\caption{Rescaled distribution function as a function of rescaled momentum for $n_0 = 0.2$ at different times with $\Q t_{\mrm{ref}} = 1500$. The original distribution function before rescaling is shown in the inset.}
	\label{fig_fp_selfsim}
\end{figure}

We start the discussion of our numerical results by presenting some properties of the well-studied attractor in non-Abelian systems far from equilibrium. Starting the (background) non-Abelian plasma at $\Q t = 0$ from the initial conditions of Eq.~(\ref{eq_init_cond}), the system quickly approaches a nonthermal fixed point, where the distribution function follows a self-similar evolution
\begin{align}
 f(t,p) = (\Q t)^\alpha f_S((\Q t)^\beta p)\,,
\end{align}
with exponents $\alpha = -4/7$ and $\beta = -1/7$. The self-similar behavior is shown in Fig.~\ref{fig_fp_selfsim}. The occupation numbers rescaled by $t^{-\alpha} f$ are plotted as functions of the rescaled momentum $t^\beta p$. Since these curves for different times fall on top of each other, this shows that the system follows a self-similar evolution, and the stationary curve corresponds to the scaling function $f_S(p)$.\footnote{The self-similar evolution with the cited values for $\alpha$, $\beta$ has been established in Refs.~\cite{Berges:2008mr,Kurkela:2011ti,Kurkela:2012hp,Berges:2012ev,Schlichting:2012es,Berges:2013fga,York:2014wja}. These values have also been extracted there from gauge-invariant observables.} 
For the simulated time in classical simulations, one observes $f_S(p) \sim (p/\Q)^{-\kappa}$ with an exponent $\kappa \approx 1.3$ \cite{Berges:2008mr,Kurkela:2012hp} while a comparison with kinetic theory predicts that it should eventually approach $\kappa = 1$ at late times \cite{York:2014wja}.

\begin{figure}[t]
	\centering
	\includegraphics[scale=0.7]{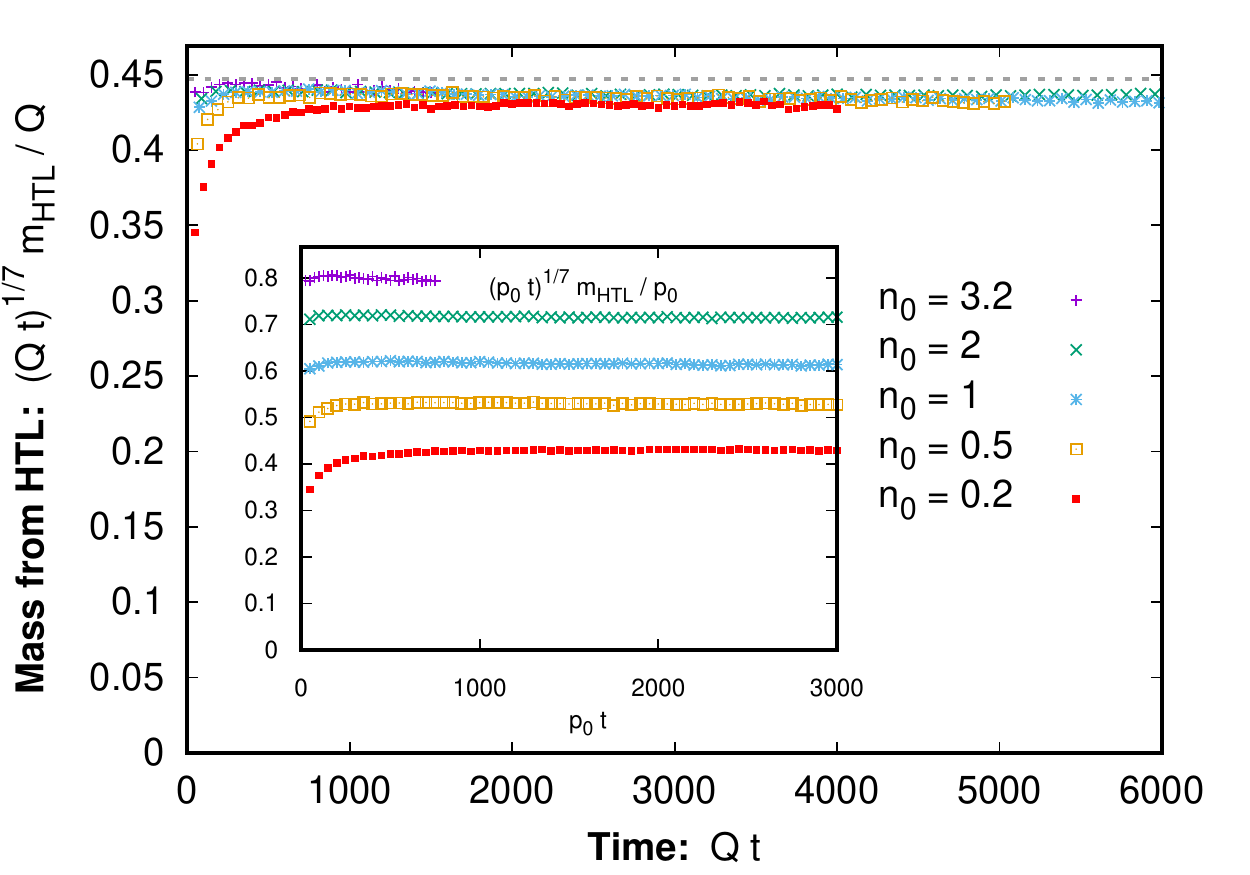}
	\caption{The mass $m_\HTL$ computed by Eq.~(\ref{eq_mass_formula}) as a function of time for different initial amplitudes $n_0$. The main figure shows its evolution rescaled by $t^{1/7}$ in units of $\Q$ while the inset shows the same in units of $\pInit$.}
	\label{fig_msqr_diff_n0}
\end{figure}

\begin{figure}[t]
	\centering
	\includegraphics[scale=0.7]{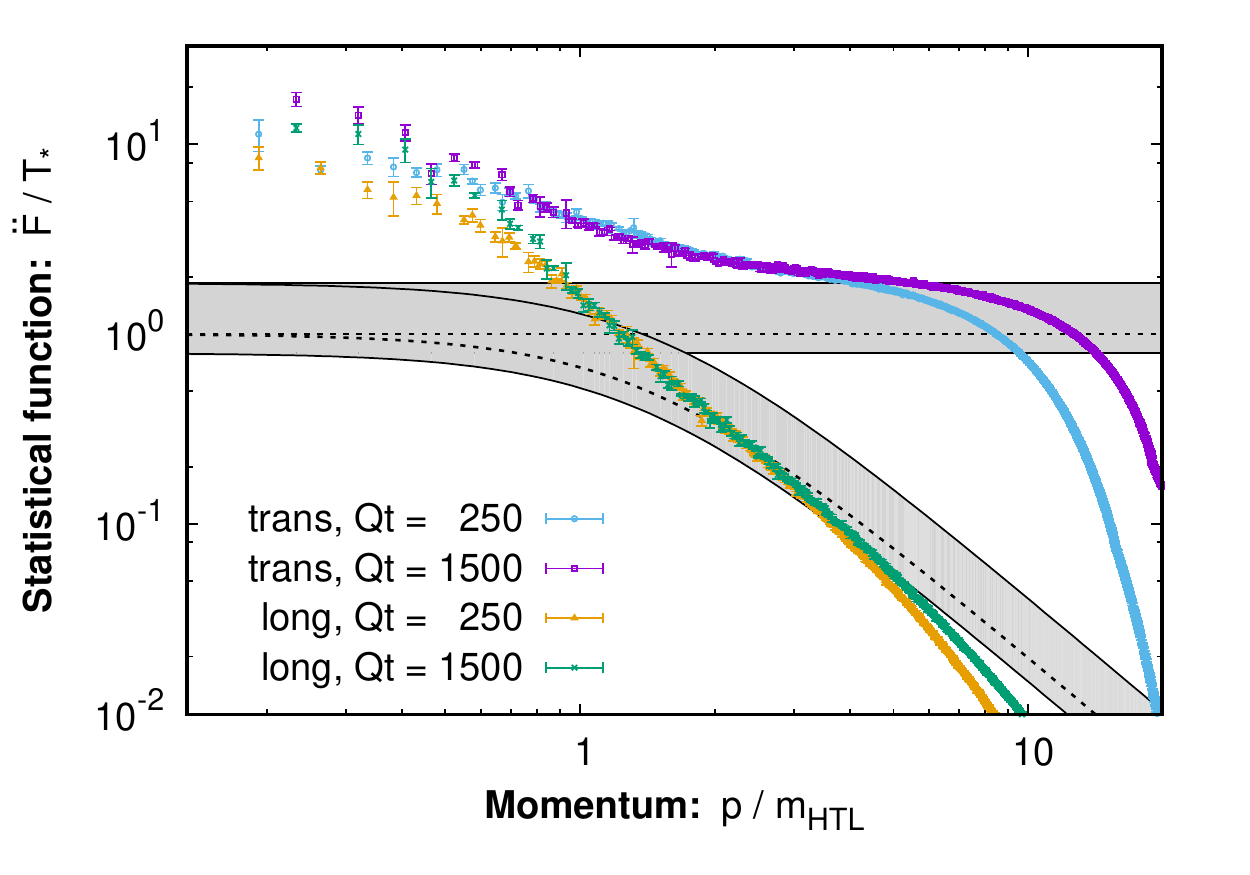}
	\caption{Shown are the transverse and longitudinal statistical correlation functions at equal times $\ddot{F}(t, \Delta t = 0, p)$ at $\Q t = 250$ and $1500$ on a large $256^3$ lattice with $\Q a_s = 0.7$ to have a good resolution of low momentum modes. For comparison, the expected HTL curves \eqref{eq_fluct_diss_equTime_HTL} are shown as gray bands the error of which results from varying the definition of the distribution function in \eqref{eq_distr_def}. Correlation functions and momenta are normalized by $T_*$ and $m_\HTL$, respectively, computed with $f_{\mrm{EE}}$.}
	\label{fig_ET_vs_EL_corrs}
\end{figure}

The physical interpretation of the self-similar evolution is a direct energy cascade, where energy density $\epsilon \sim \int \mrm d^3 p\; \omega(p) f(t,p) \sim \Lambda^4 f(\Lambda) = \mrm{const}$ is conserved and transported to higher momenta as $\Lambda / \Q \sim (\Q t)^{1/7}$, with $\Lambda$ being the hard momentum scale that dominates the energy density. Indeed, the evolution in this time regime only depends on the energy density \cite{York:2014wja}, while details of the initial conditions are washed away by the transient evolution to the nonthermal fixed point \cite{Kurkela:2012hp,Berges:2013lsa}. Since our definition of $\Q$ relies on the energy density, all quantities should become independent of the initial conditions at late times $\Q t$. 

This is demonstrated at the example of the HTL mass of \eqref{eq_mass_formula}. Because of the parametric relation $m_\HTL^2 \sim \int \ud^3 p\, f(t,p) / p$, the self-similar evolution of $f(t,p)$ should lead to $m_\HTL/\Q \sim (\Q t)^{-1/7}$. We show its evolution in Fig.~\ref{fig_msqr_diff_n0} for different initial amplitudes $n_0$. One observes that the mass indeed becomes proportional to the expected power law behavior for all amplitudes at sufficiently late times. When expressed in units of $\Q$, the masses for different $n_0$ are seen to even collapse to a single curve at late times, which signals that the dynamics becomes insensitive to the initial conditions. This is already the case at $\Q\tpert = 1500$ for $n_0 = 0.2$, which is the set of parameters we will be using in the following. For comparison, we show $m_\HTL^2 / \pInit^2$ as a function of $\pInit t$ in the inset, where the curves stay clearly apart and do not coincide. 

The different power laws of the mass and hard scale lead to an increasing scale separation with time $m_\HTL / \Lambda \sim (\Q t)^{-2/7}$. Since a scale separation is one of the main assumptions in the HTL formalism that was discussed in \se\ref{sec_HTL_theory}, one expects this formalism to be applicable to the considered highly occupied system. In the following, we will show our results and compare them to the respective HTL predictions at LO in $m_\HTL / \Lambda$. In principle, we could also tune $m_\HTL / \Lambda$ by extending simulations to late times to measure subleading effects. However, computational costs restrict such an analysis since the required lattices would also need to increase to prevent lattice artifacts.

\begin{figure}[t]
	\centering
	\includegraphics[scale=0.7]{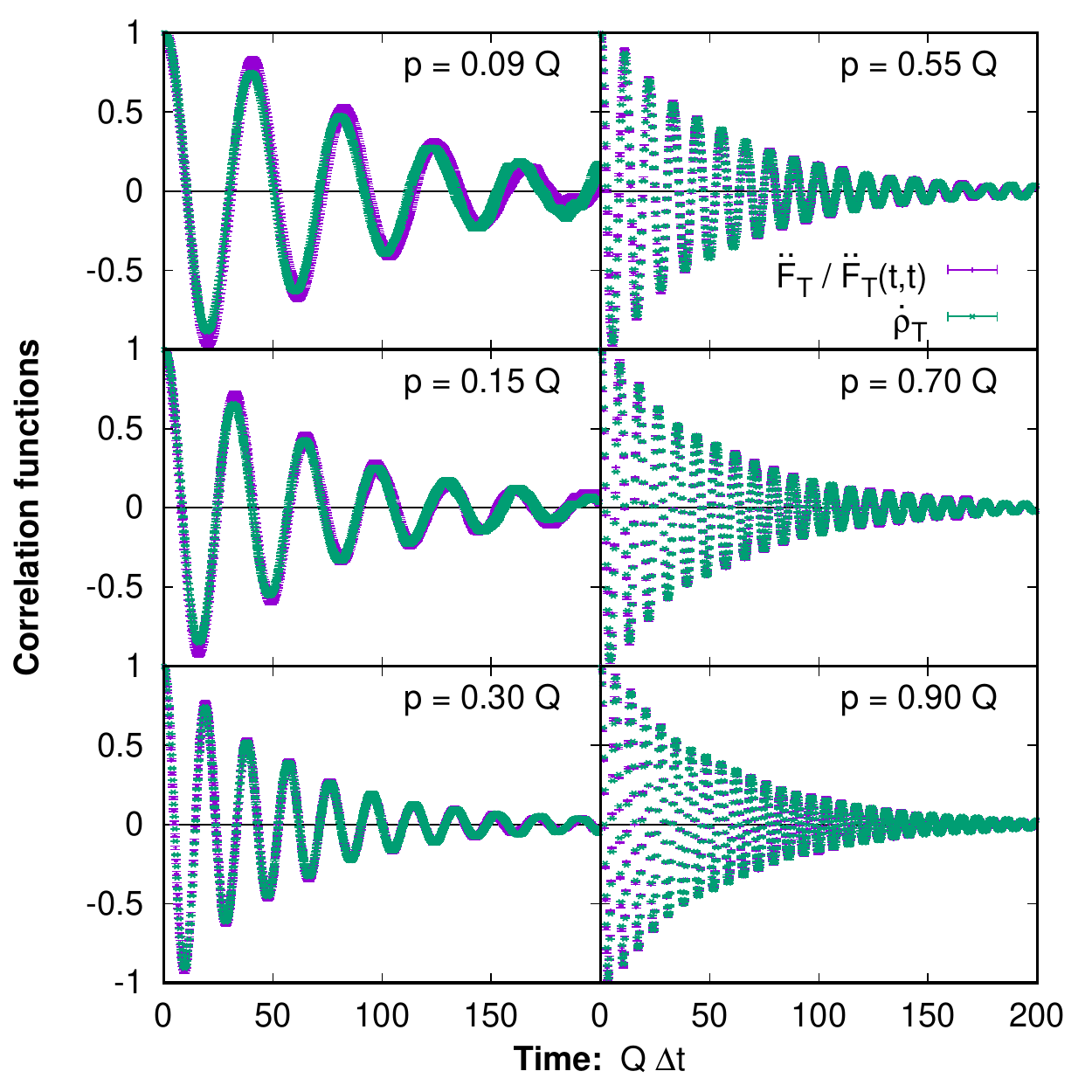}
	\caption{Statistical and spectral functions as functions of time $\Delta t$ for different momenta. While $\dot{\rho}_T$ always starts at $1$ for $\Delta t = 0$, we divide $\ddot{F}_T$ by the equal-time correlation function $\ddot{F}_T(t,\Delta t = 0,p)$ to normalize it accordingly.}
	\label{fig_rho_F_dt_multi}
\end{figure}

\begin{figure}[t]
	\centering
	\includegraphics[scale=0.7]{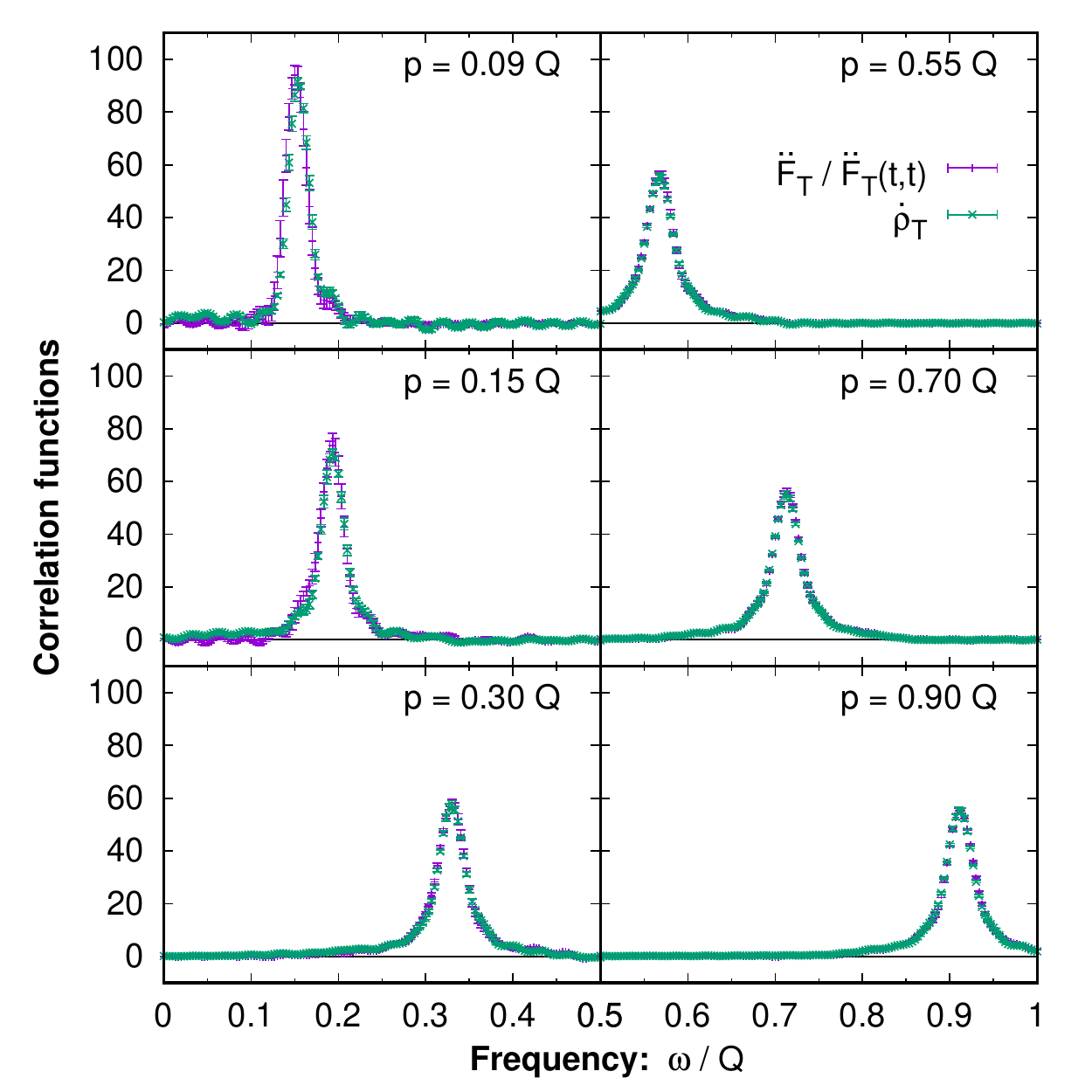}
	\caption{Statistical and spectral functions of \fig\ref{fig_rho_F_dt_multi} Fourier transformed to frequency space, where $\Q \dtmax = 200$ was chosen.}
	\label{fig_rho_F_w_multi}
\end{figure}

\subsection{Comparing spectral and statistical correlation functions}
\label{sec_comp_spec_stat}

We start our discussion of spectral and statistical correlation functions by studying their equal-time correlations $\ddot{F}(t, \Delta t = 0, p)$ and $\dot{\rho}(t, \Delta t = 0, p)$, the latter of which is fixed by \eqref{eq_rho_sum_rules}. In Fig.~\ref{fig_ET_vs_EL_corrs} we show the transverse and longitudinal correlation functions $\ddot{F}_{T/L} (t, \Delta t = 0, p)$ at times $\Q t = 250$ and $1500$. Gray areas show the HTL prediction in \eqref{eq_fluct_diss_equTime_HTL} with error bands. Instead of following a straight line, as predicted by HTL, the transverse correlation depends on momentum, following an approximate power law $T_*\,(p/\Q)^{-0.3}$ between $m_\HTL$ and $\Lambda$ and surpassing the gray band denoting $T_*$ at low momenta. This power law is connected to the $f \sim (p/\Q)^{-\kappa}$ behavior with $\kappa > 0$ discussed in \se\ref{sec_selfsim_attr} while $\kappa = 1$ is predicted in HTL for a perfect scale separation.  

Similarly, the longitudinal correlation shows deviations from the expected $T_*\, \dot{\rho}_L(t, \Delta t = 0, p)$ evolution. At momenta below the mass $p \lesssim m_\HTL$, both $\ddot{F}_T$ and $\ddot{F}_L$ are enhanced and approximately coincide. At high momenta $p \gtrsim \Lambda$ they decrease exponentially, which is beyond the HTL prediction since the latter is only expected to hold for $p \ll \Lambda$. As time proceeds, the longitudinal correlation $\ddot{F}_L$ is observed to approach the lower bound of the HTL predicted gray band while the transverse correlation function $\ddot{F}_T$ typically comes closer to the upper bound of the HTL prediction. 

The HTL relation between the equal-time correlation functions \eqref{eq_fluct_diss_equTime_HTL} that we were testing is based on the more general fluctuation-dissipation relation \eqref{eq_fluct_diss_rel_HTL} that predicts the same connection $T_*$ between $\ddot{F}$ and $\dot{\rho}$ as functions of relative time $\Delta t$ and of $\omega$. Focusing here on the transverse polarization, we will therefore study the relation between $\ddot{F}_T$ and $\dot{\rho}_T$ in the considered system. The normalization of the spectral function is fixed by the sum rule \eqref{eq_rho_sum_rules}. Similarly, $\ddot{F}_T(\tcent,\Delta t,p) / \ddot{F}_T(\tcent,\Delta t = 0,p)$ satisfies the same normalization relation. Their evolution is shown in \fig\ref{fig_rho_F_dt_multi} as functions of relative time $\Delta t$ at time $\Q\tpert = 1500$. One observes that they lie on top of each other to good approximation and show damped oscillations. Their corresponding Fourier transforms are presented in \fig\ref{fig_rho_F_w_multi}. They are seen to also coincide in frequency space, establishing the relation
\begin{align}
 \label{eq_ddF_drho_rel}
 \ddot{F}_T(\tcent,\omega,p) \approx \ddot{F}_T(\tcent,\Delta t = 0,p) \, \dot{\rho}_T(\tcent,\omega,p), 
\end{align}
and equivalently as functions of $\Delta t$. This relation corresponds to a generalized fluctuation-dissipation relation for our far-from-equilibrium situation by constraining the connection between the correlation functions not to depend on frequency or on the relative time. Compared to the HTL relation \eqref{eq_fluct_diss_rel_HTL} at LO, this translates to allowing $T_*(\tcent) \mapsto \ddot{F}_T(\tcent,\Delta t = 0,p)$ to depend on time and momentum, which is consistent with what we observed in \fig\ref{fig_ET_vs_EL_corrs}. 

These observations show that the correlation function $\ddot{F}_{T}(\tcent,\omega,p)$ in the considered highly occupied system is similar to its LO HTL expectation by satisfying a generalized fluctuation-dissipation relation but shows deviations for its amplitude $\ddot{F}_T(\tcent,\Delta t = 0,p)$ that becomes momentum dependent instead of being equal to $T_*$. One of the reasons for this is that $m_\HTL / \Lambda$ is not negligibly small, which affects the form of $f(t,p)$ and thus, the form of $\ddot{F}_T(\tcent,\Delta t = 0,p)$ and the values of $m_\HTL$ and $T_*$. We also observe deviations for momenta below $m_\HTL$ that cannot be explained by the HTL expressions at hand even qualitatively and may be influenced by processes at the magnetic scale.

\begin{figure}[t]
	\centering
	\includegraphics[scale=0.7]{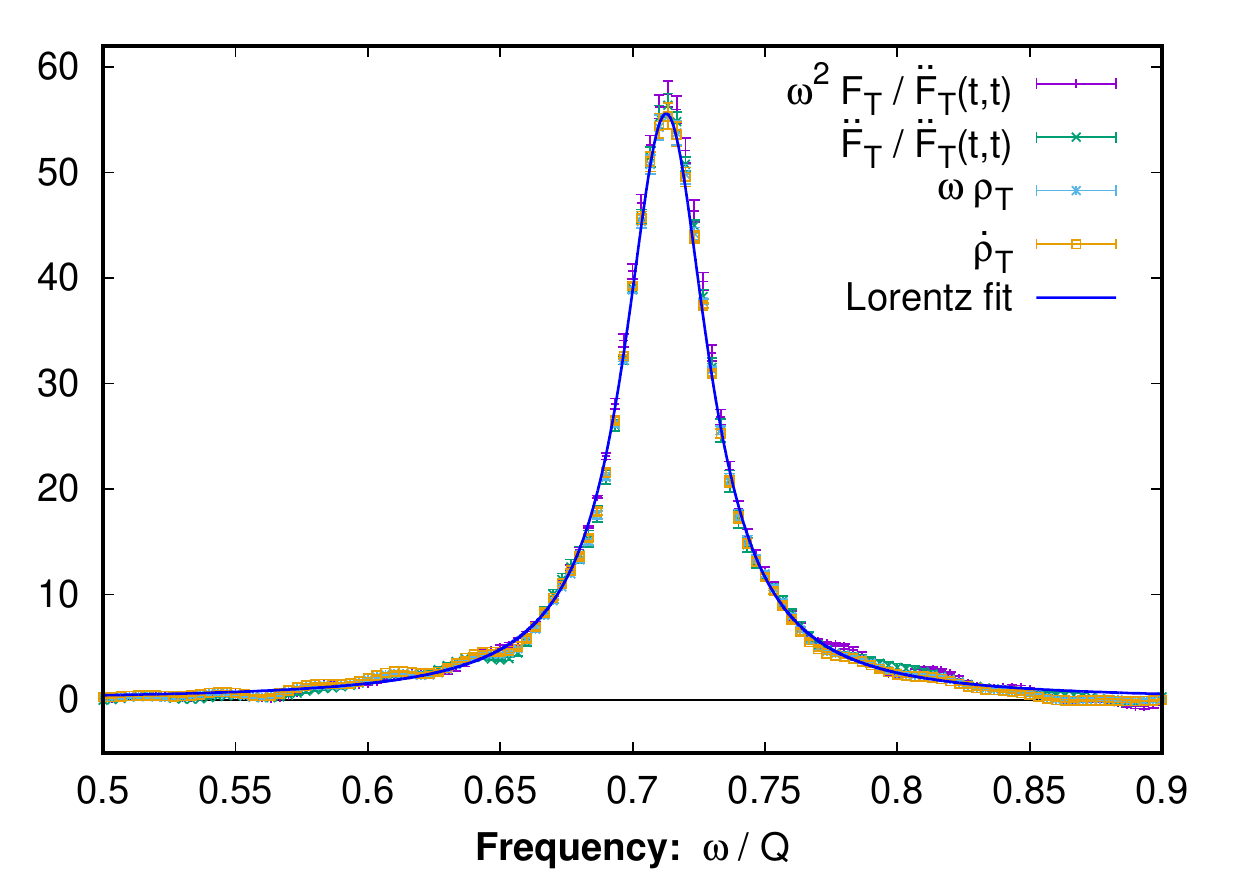}
	\caption{Transverse statistical and spectral functions in frequency space for $p = 0.7\, \Q$. A Lorentzian fit function \eqref{eq_lor_curve} is additionally shown for comparison.}
	\label{fig_lorentz}
\end{figure}

Moreover, the spectrum in \fig\ref{fig_rho_F_w_multi} includes a narrow peak for each momentum, which follows a Lorentzian form\footnote{The small deviations from this form at low momenta like $p = 0.09\, \Q$ can be attributed to the finite time window employed and are expected to vanish with a larger time window for the Fourier transform.}
\begin{align}
 \label{eq_lor_curve}
 g_{\mrm{Lor}}(\omega) = \frac{A}{\pi}\, \frac{\gamma_T}{(\omega - \omega_T)^2 + \gamma_T^2}\,.
\end{align}
Here $A$ is a normalization constant and $\omega_T$ and $\gamma_T$ are the transverse dispersion relation and damping rate, respectively, all of which can in general depend on momentum and time. This shape is demonstrated in \fig\ref{fig_lorentz} at the example of $p = 0.7\, \Q$. To the correlation functions shown in \fig\ref{fig_rho_F_w_multi} we added $\omega^2 F_T / \ddot{F}_T(t,\Delta t = 0,p)$ and $\omega \rho_T$, which confirm that time derivatives become frequency factors in frequency space despite the residual central time dependence. Most importantly, their shape indeed matches with \eqref{eq_lor_curve}, which establishes the existence of quasiparticles in the considered system.\footnote{Note that because of $\omega_T \gg \gamma_T$, the spectral function $\rho_T$ also follows a Lorentzian form in frequency space with approximately the same $\omega_T$ and $\gamma_T$. We have checked this explicitly.} Their dispersion relation and damping rate are discussed in the following subsection.

\subsection{Dispersion relation, damping rate and Landau cut}
\label{sec_HTL_disp_damp}

Proceeding with our discussion of correlation functions at unequal time, we study the transverse dispersion relation $\omega_T$ of the observed quasiparticle peak and compare it to the expected HTL curve at LO $\omega_T^\HTL$ discussed in \se\ref{sec_HTL_theory}. While the latter involves the plasmon frequency $\wplas^\HTL$ and asymptotic mass $m_\HTL$ given by \eqref{eq_wplas_m_rel} and \eqref{eq_mass_formula}, in general, the plasmon frequency is defined as the frequency of the zero mode, 
\begin{align}
 \label{eq_wplas_general}
 \wplas = \omega(0) = \lim_{p \rightarrow 0} \omega_T(p) = \lim_{p \rightarrow 0} \omega_L(p),
\end{align}
where $\omega_L(p)$ is the longitudinal dispersion relation that will be discussed and studied in \se\ref{sec_long_corrs}. Since at $p = 0$ there is no distinction between transverse and longitudinal polarizations, both dispersion relations have to coincide. Similarly, the asymptotic mass is defined as the mass in the relativistic dispersion relation $\sqrt{m^2 + p^2}$ that is expected to be seen for high momenta $p \gg m$. One way of expressing this relation is 
\begin{align}
 \label{eq_infty_m_general}
 m = \lim_{p \rightarrow \infty} \sqrt{\omega_T^2 - p^2}.
\end{align}
Both definitions are consistent with the Taylor expanded LO HTL dispersion relation in \eqref{eq_wHTL_Taylor}. However, they also enable us to measure $\wplas$ and $m$ independently and to compare their values to the HTL predictions.

\begin{figure}[t]
	\centering
	\includegraphics[scale=0.7]{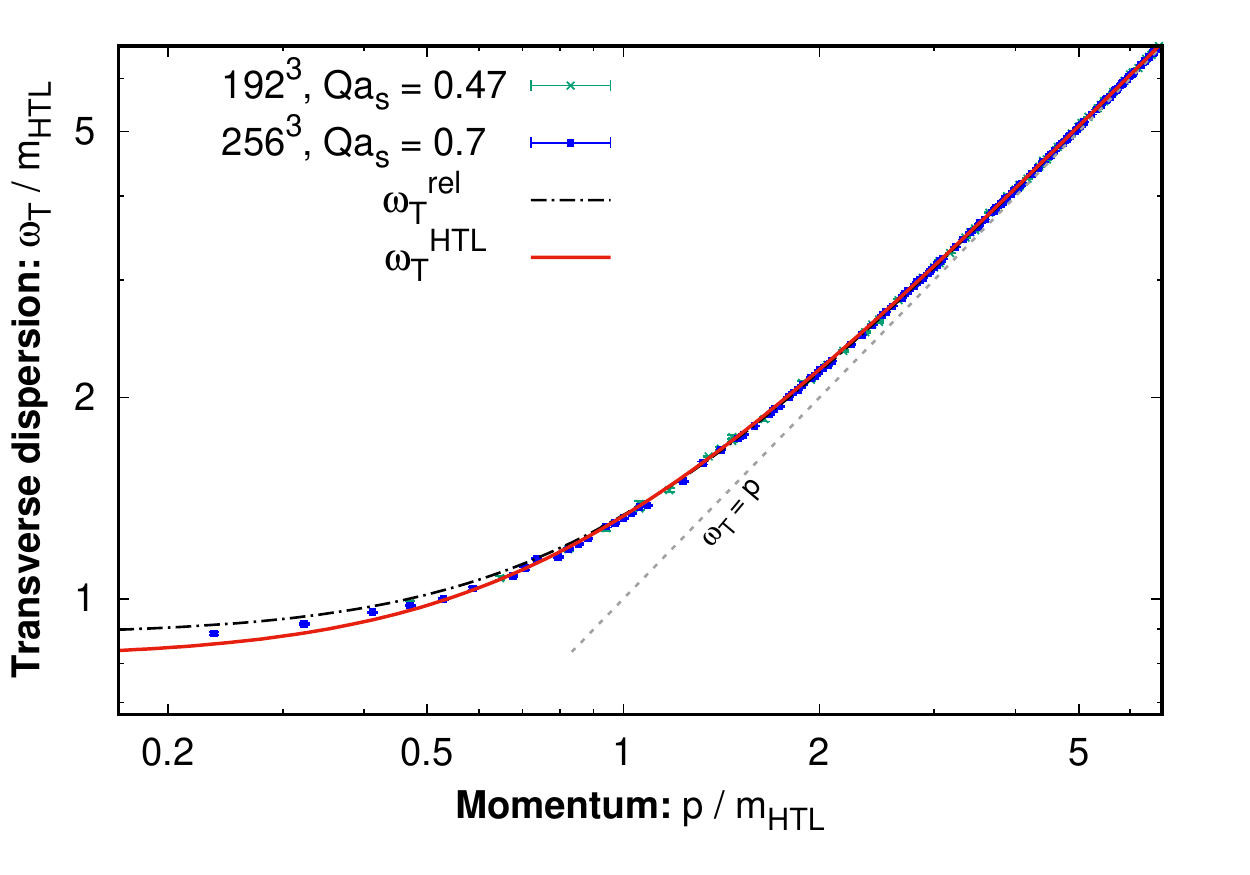}
	\includegraphics[scale=0.7]{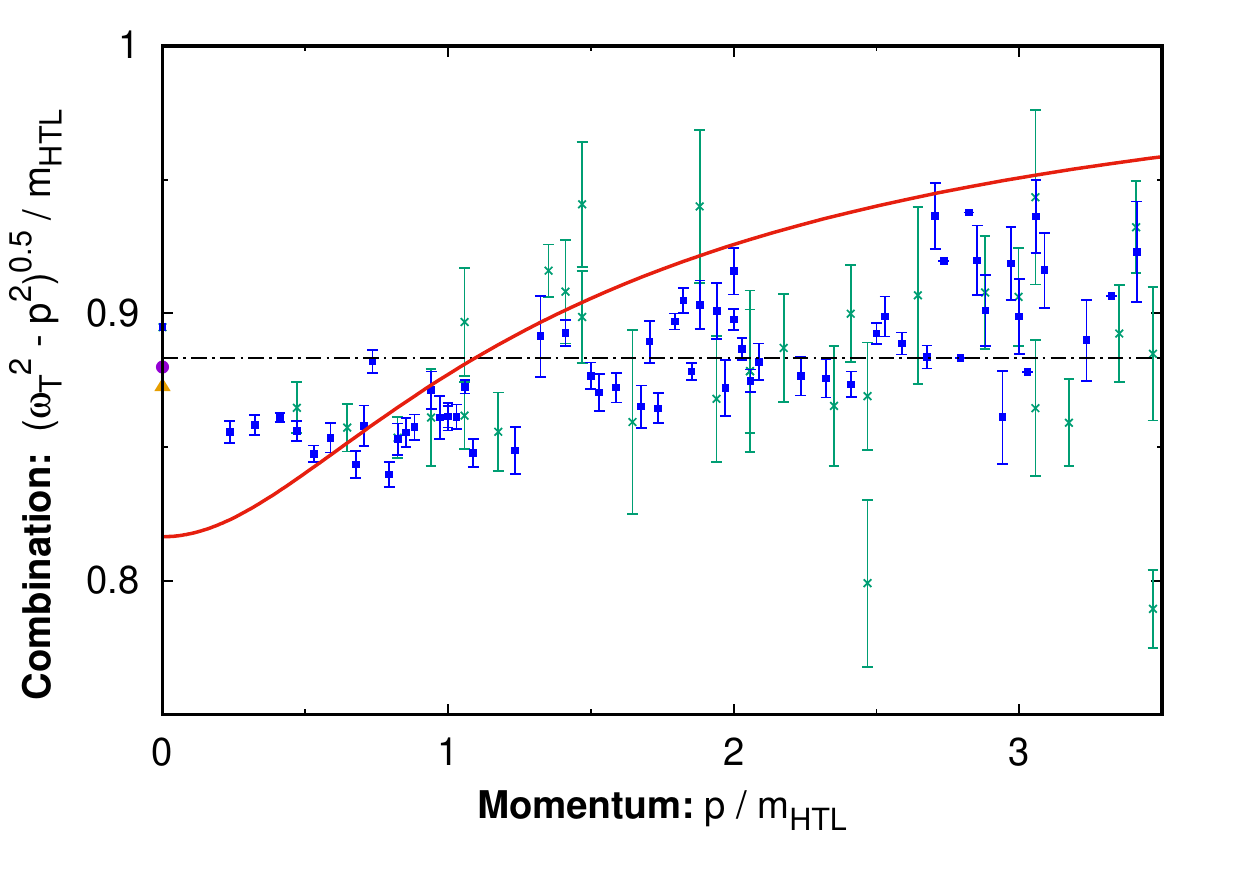}
	\caption{{\bf Upper:} Transverse dispersion relation $\omega_T(p)$ deduced from $\dot{\rho}_T(\tpert,\omega,p)$ by finding the location of the maximum of the quasiparticle peak for each momentum in a logarithmic plot. The dashed black curve depicts a fit to a relativistic dispersion $\omega_T^\rel$ while the red solid curve shows the expected dispersion relation from HTL calculations at LO with HTL mass $m_\HTL = 0.149 \, \Q$. For comparison, we also show the ultra-relativistic dispersion relation $\omega_T = p$ as a gray dashed curve. 
	{\bf Lower:} The expression $\sqrt{\omega_T^2 - p^2}$ is shown for the same data in a linear plot. We also included the zero mode frequency $\omega(0)$ from the $\Q a_s = 0.7$ and $\Q a_s = 0.47$ discretizations of the longitudinally polarized systems in \fig\ref{fig_disprel_long} by an orange triangle and purple circle, respectively.}
	\label{fig_disprel_trans}
\end{figure}

More generally, we can now compare our results to the expected LO curves from HTL. We start with the transverse dispersion relation $\omega_{T}(p)$ in the upper panel of \fig\ref{fig_disprel_trans}. It is deduced from $\dot{\rho}_T(\tpert,\omega,p)$ by finding the frequency of the maximum of the quasiparticle peak for each momentum, but we have also checked that the dispersion resulting from the maxima of $\rho_T(\tpert,\omega,p)$ is consistent within uncertainties. One observes that the extracted dispersion relation does not show sensitivity to the different discretization parameters employed, which indicates that this observable is close to its continuum limit. This can even be observed at low momenta. For instance, the smallest momentum of the $\Q a_s = 0.47$ system is located at $p \approx 0.48\, m_\HTL$ but it is barely visible because the corresponding point of the $\Q a_s = 0.7$ data lies almost on top of it. 

For comparison, we included a relativistic dispersion relation
\begin{align}
 \omega_T^\rel = \sqrt{m_\rel^2 + p^2}, 
\end{align}
with a fitted mass value $m_\rel = 0.132\, \Q$ into the upper panel of \fig\ref{fig_disprel_trans} . We also show there the numerically computed transverse HTL dispersion relation $\omega_{T}^{\HTL}$, where we use the mass parameter $m_\HTL = 0.149 \, \Q$ of Eq.~\eqref{eq_mHTL_value}, which comes from a computation within the HTL formalism with the distribution function $f_{\mrm{EE}}$ while other definitions lead to deviating values. Interestingly, both functional forms provide good overall descriptions of the data. For the HTL curve, we get the same value for $m_\HTL$ when fitting $\omega_{T}^{\HTL}$ to our data. This justifies a posteriori our choice of the mass value for $m_\HTL$ in our previous and following figures where data is compared to HTL predicted curves. 

On the other hand, there are discrepancies at low and high momenta. They are better visible in the lower panel of \fig\ref{fig_disprel_trans} where we show $\sqrt{\omega_T^2 - p^2}$ as a function of momentum for the curves of the upper panel. While the data points from this combination are quite noisy, one observes some systematic behavior. If the transverse dispersion relation followed a simple relativistic form $\omega_T^\rel$, the data points would be constant and equal to $m_\rel$. Instead, one observes that apart from the zero mode this combination steadily grows overshooting this mass value at high momenta and being smaller at low momenta. This is qualitatively similar to the behavior of the HTL curve but the data points are systematically lower at high momenta and higher at low momenta. This systematics can be also observed in the upper panel of \fig\ref{fig_disprel_trans} when compared to the HTL curve, i.e., frequencies are shifted to slightly larger values at low momenta. Moreover, we will see in \se\ref{sec_long_corrs} that the longitudinal dispersion has the same systematic behavior when compared to the corresponding HTL curve. 

After this qualitative discussion of the functional form of the extracted transverse dispersion relation, we can deduce the values for the plasmon frequency and the asymptotic mass from our data. We measure the plasmon frequency directly by finding the maximum in frequency space of the spectral function at $p = 0$
\begin{align}
 \label{eq_wplas_data}
 \wplas^\fit(\Q t = 1500) / \Q = 0.132 \pm 0.002. 
\end{align}
The corresponding points at $p = 0$ are depicted in the lower panel of \fig\ref{fig_disprel_trans}. They correspond to different discretizations, which lead to the same value within the given uncertainty. This value is larger than the predicted value $\wplas^\HTL = 0.122\,\Q$ but agrees with the value from the relativistic dispersion fit $\wplas^\rel = m_\rel$. 

The asymptotic mass $m$ is deduced in our simulations by fitting $\omega_{T}^{\HTL}(p)$ to the observed dispersion relation $\omega_T(p)$ for high momenta $p_{\mrm{min}} \leq p \leq 1\,\Q = 6.7\, m_\HTL$. In practice, we first compute the HTL dispersion relation at various momenta $p$, interpolate the solution and fit the interpolation function to the data. Varying the lower momentum between $p_{\mrm{min}} = 2\, m_\HTL - 4\, m_\HTL$ and considering the maxima of both $\rho_T$ and $\dot{\rho}_T$ provides an estimate for the systematic error while the $(\mrm{fit})$ value is the typical error of the fit
\begin{align}
 \label{eq_extracted_mass}
 m_\fit / \Q = 0.138 \pm 0.002\; (\mrm{sys}) \pm 0.0015\; (\mrm{fit}). 
\end{align}
The fitted value for the asymptotic mass is smaller than the value used for $m_\HTL$, which is consistent with the lower panel of \fig\ref{fig_disprel_trans} because the $\sqrt{\omega_T^2 - p^2}$ data points lie below the corresponding HTL curve. According to \eqref{eq_wHTL_Taylor}, $\omega_T^\rel$ approximates the HTL curve $\omega_{T}^{\HTL}$ at large momenta. However, the value from the relativistic dispersion fit $m_\rel$ is lower than the extracted value $m_\fit$, which is also consistent with the lower panel of \fig\ref{fig_disprel_trans}. 

With this, we can compute the relation between the plasmon frequency and the mass. At LO in the HTL framework one has $\wplas^\HTL / m_\HTL = \sqrt{2/3} \approx 0.8165$, while for the relativistic dispersion one has $\wplas^\rel / m_\rel = 1$. We find
\begin{align}
 \label{eq_wplas_m_ratio}
 \frac{\wplas^\fit}{m_\fit}(\Q t = 1500) = 0.957 \pm 0.028\,,
\end{align}
which is even closer to $1$ than to the HTL expected $\sqrt{2/3}$. However, we emphasize that the extracted transverse dispersion agrees well with both functional forms and shows deviations from both forms as well. On the other hand, as we will see, we will observe the HTL predictions of a Landau cut region and a different dispersion relation for longitudinally polarized modes. This, together with the agreement with the transverse dispersion relation, shows that the HTL formalism provides a good overall description of our data, while systematic deviations can be taken as an indication of effects beyond this formalism at LO.

\begin{figure}[t]
	\centering
	\includegraphics[scale=0.7]{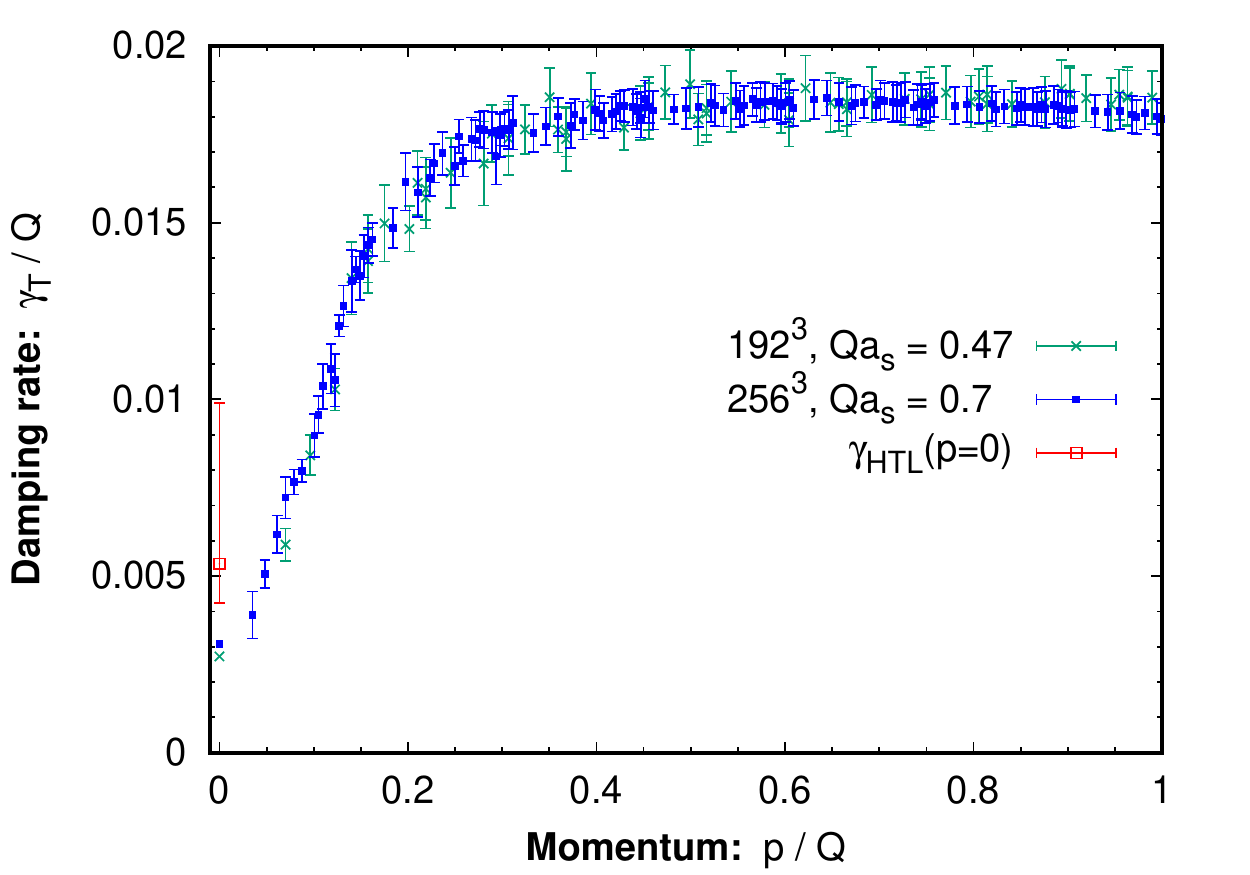}
	\caption{Transverse damping rate $\gamma_T(p)$ deduced from $\dot{\rho}_T(\tpert,\Delta t, p)$ by fitting to the form of a damped oscillator \eqref{eq_damped_oscil} within a time range of $\Q \Delta t \leq 70$. The error bars are calculated as a sum of the error of the mean and the average fitting error. The expected HTL value $\gamma_\HTL(p=0)$ is also included. For the data points at $p=0$ we fitted a damped oscillator to already averaged data within $\Q \Delta t \leq 100$, therefore not extracting error bars.}
	\label{fig_damp_rate}
\end{figure}

Let us now proceed with the discussion of the damping rate of transverse quasiparticles. Within the HTL framework, poles of the transverse retarded propagator at leading order \eqref{eq_GR_HTL} correspond to Delta functions $\delta(\omega - \omega_{T}^{\HTL}(p))$ in $\rho_T$. However, realistic quasiparticle peaks involve a finite width, which corresponds to a non-vanishing damping rate $\gamma_T(p)$, as was observed in \se\ref{sec_comp_spec_stat}. This is an effect of subleading order in the HTL framework and may contain non-perturbative contributions. So far, the damping rate within HTL has only been calculated at $p = 0$ \cite{Braaten:1990it}. Our employed numerical framework provides a unique opportunity to access the momentum dependence of this quantity out of equilibrium. 

Instead of working in Fourier space, one can measure $\gamma_T(p)$, as well as $\omega_T(p)$, also directly in the time domain $\Delta t$. A Lorentzian peak Fourier transforms to a damped oscillator, which reads
\begin{align}
 \label{eq_damped_oscil}
 g_{\mrm{d.osc.}}(\Delta t) = e^{-\gamma_T\, \Delta t}\, \cos(\omega_T\, \Delta t),
\end{align}
at the example of $\dot{\rho}_T$. Using this fitting curve directly in the time domain, we have checked that the dispersion relation $\omega_T(p)$ extracted in this way coincides with the data shown in \fig\ref{fig_disprel_trans}. However, it turns out that this procedure provides more accurate values for the damping rate than the corresponding measurement in frequency space, where in practice, the finite time window for the Fourier transform leads to deviations from Lorentzian peaks especially at low momenta, introducing additional uncertainties on the damping rates. 

The damping rate extracted using \eqref{eq_damped_oscil} is shown in \fig\ref{fig_damp_rate} for different discretizations that lie on top of each other within uncertainties. The data points are obtained by averaging over the damping rate of different simulations while the error bars are a sum of the error of the mean and the average fitting error. For low momenta $p \lesssim 0.15\,\Q$ (i.e., $p \lesssim m_\HTL$), the damping rate is observed to increase before it eventually flattens at higher momenta. We have also included the expected value at $p = 0$ within the HTL formalism that has been computed in Ref.~\cite{Braaten:1990it} 
\begin{align}
 \label{eq_gamma_0}
 \gamma_\HTL(0) = 6.63538\, \frac{g^2 N_c T_*}{24\pi}\,,
\end{align}
where we have replaced $T \mapsto T_*$ in the original formula and estimated the error bars by employing different definitions of the distribution function in \eqref{eq_distr_def}. Our data is roughly consistent with the HTL predicted value. The extracted damping rate $\gamma_T(p)$ in \fig\ref{fig_damp_rate} is one of the main results of this work.

\begin{figure}[t]
	\centering
	\includegraphics[scale=0.7]{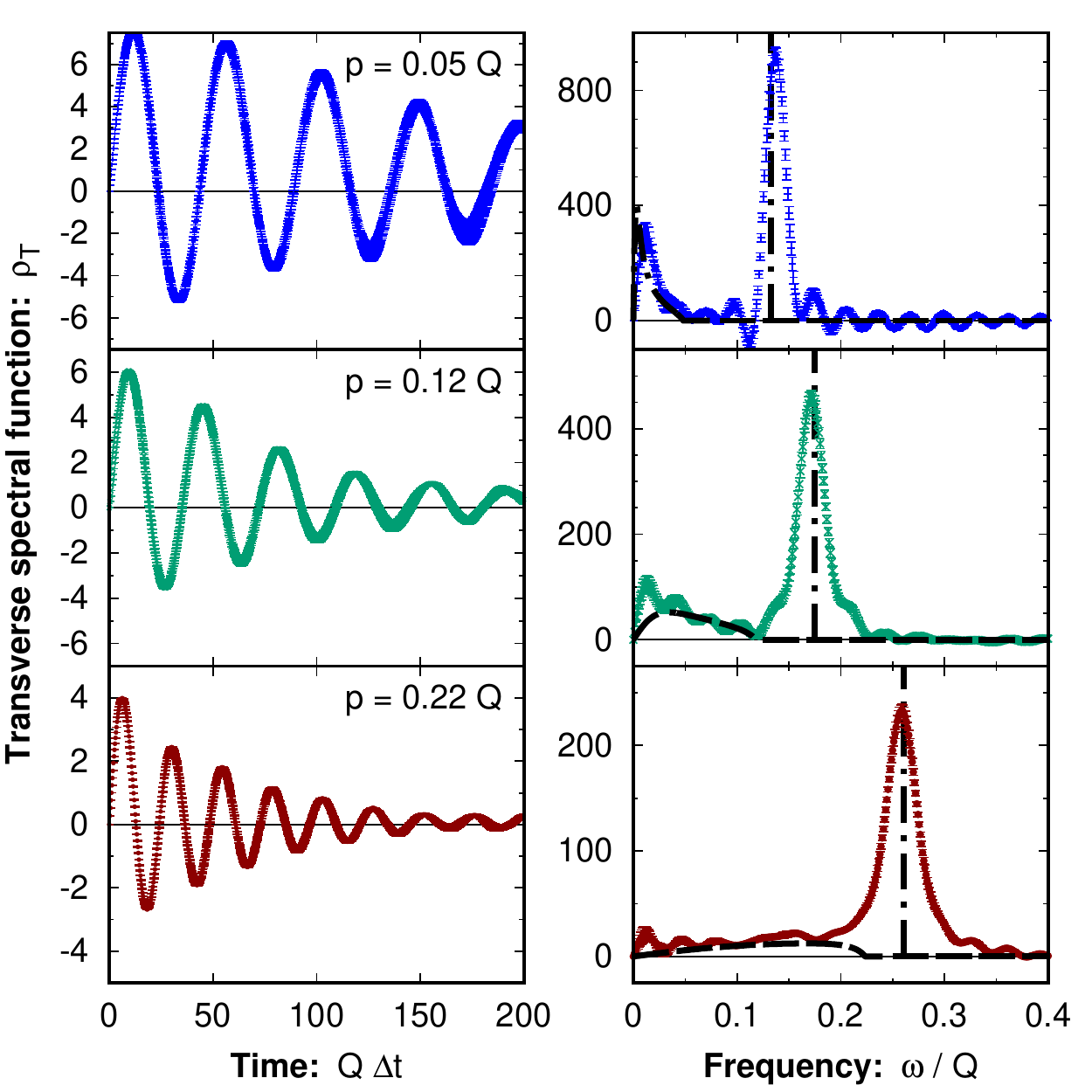}
	\caption{The spectral function $\rho_T$ is shown in both time domain and frequency domain for different low momenta $p$, where $\Q \dtmax = 200$ has been used for the Fourier transform. Dashed black lines show the expected HTL curves given by \eqref{eq_rhoT_HTL} with the HTL mass $m_\HTL = 0.149 \, \Q$.}
	\label{fig_Landau_cut}
\end{figure}

In our considerations, we have only talked about the dominant quasiparticle peak so far. However, the transverse spectral function has a much richer structure, also involving the Landau cut as discussed in \se\ref{sec_HTL_theory}. Our results for the spectral function $\rho_T$ are shown in \fig\ref{fig_Landau_cut} for different momenta, where we also included the expected HTL curves. While the additional $\omega$-factor suppressed the Landau cut region in \fig\ref{fig_rho_F_w_multi}, the corresponding structure at low frequencies for $\omega \leq p$ is clearly visible in \fig\ref{fig_Landau_cut}. Moreover, it agrees well with the HTL curves at LO. Note that the only parameter in these curves is the mass $m$, for which we employ $m_\HTL$ to be consistent with \fig\ref{fig_disprel_trans}. Therefore, no free parameter is left and the agreement with the curves indicates that the HTL framework provides a valid description for the Landau cut region even far from equilibrium. 

Small discrepancies at the lowest frequencies $\omega \lesssim 0.02\,\Q$ can be attributed to the finite resolution in frequency space due to limitations of the time window for the Fourier transform. On the other hand, the quasiparticle peak of low momenta like $p = 0.05\, \Q$ is located at a slightly larger frequency than expected at LO in HTL. This behavior is consistent with our observation of the dispersion relation in \fig\ref{fig_disprel_trans}.

\begin{figure}[t]
	\centering
	\includegraphics[scale=0.7]{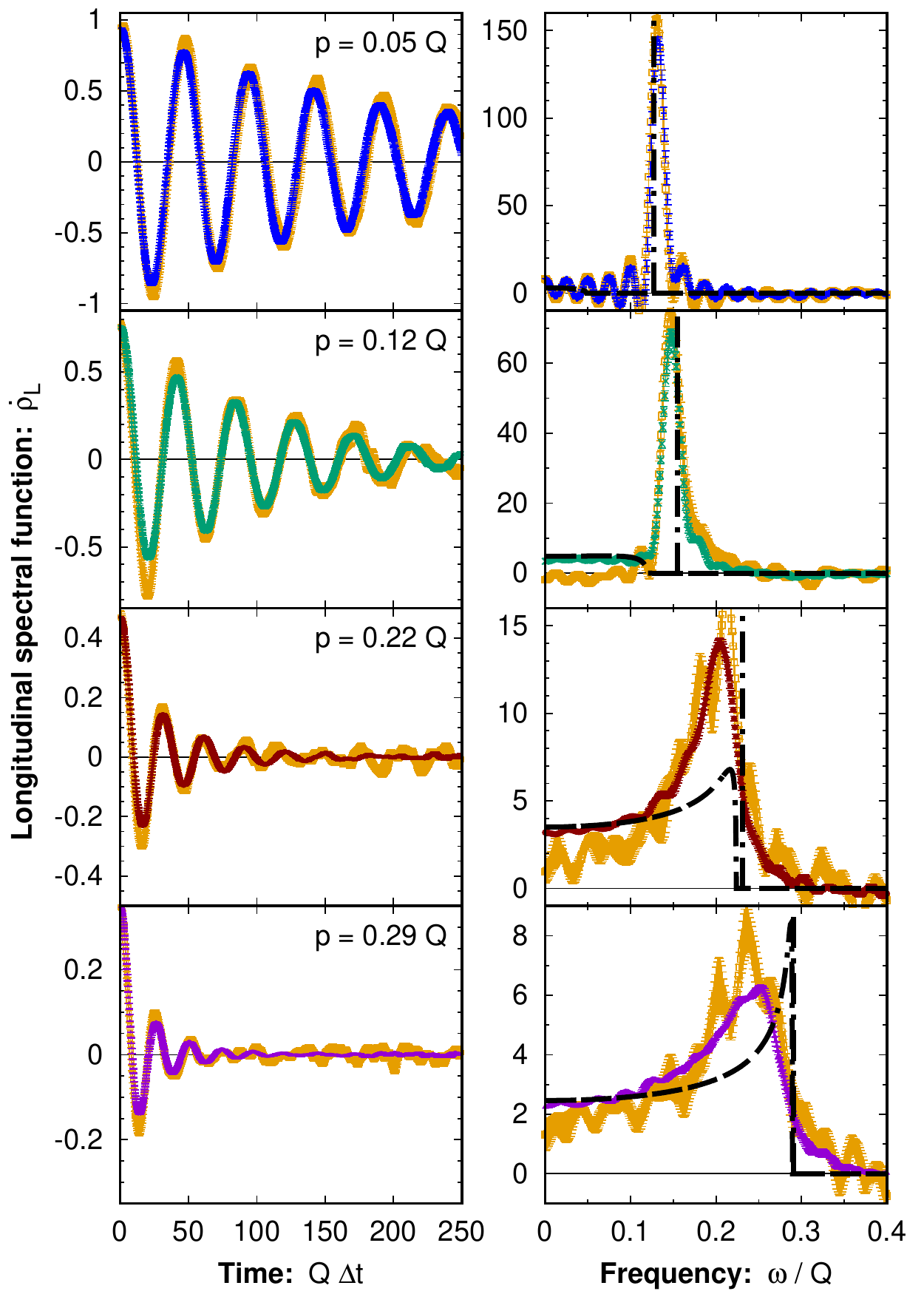}
	\caption{The longitudinal spectral function $\dot{\rho}_L$ is shown in both time domain and frequency domain for different momenta $p$, where $\Q \dtmax = 250$ has been used for the Fourier transform. Our data for the normalized longitudinal statistical correlation function $\ddot{F}_L / \ddot{F}_L(t,\Delta t = 0,p)\; \dot{\rho}_L^\HTL(t,\Delta t = 0,p)$ is depicted by orange curves. Dashed black lines show the longitudinal HTL curve $\dot{\rho}_L^\HTL$ of \se\ref{sec_HTL_theory}, which corresponds to a sum of the Landau cut region and the quasiparticle peak.}
	\label{fig_rho_dt_w_long}
\end{figure}

\begin{figure}[t]
	\centering
	\includegraphics[scale=0.7]{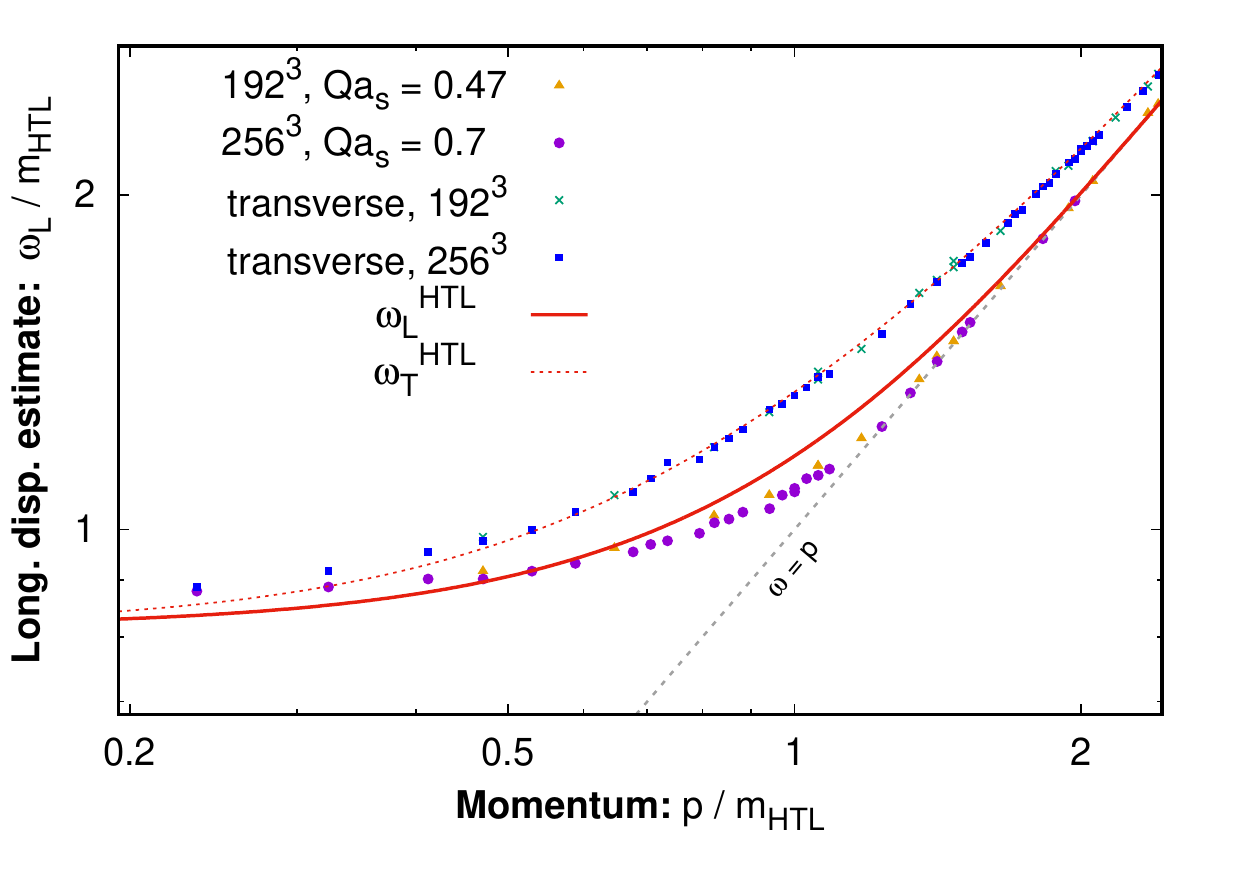}
	\includegraphics[scale=0.7]{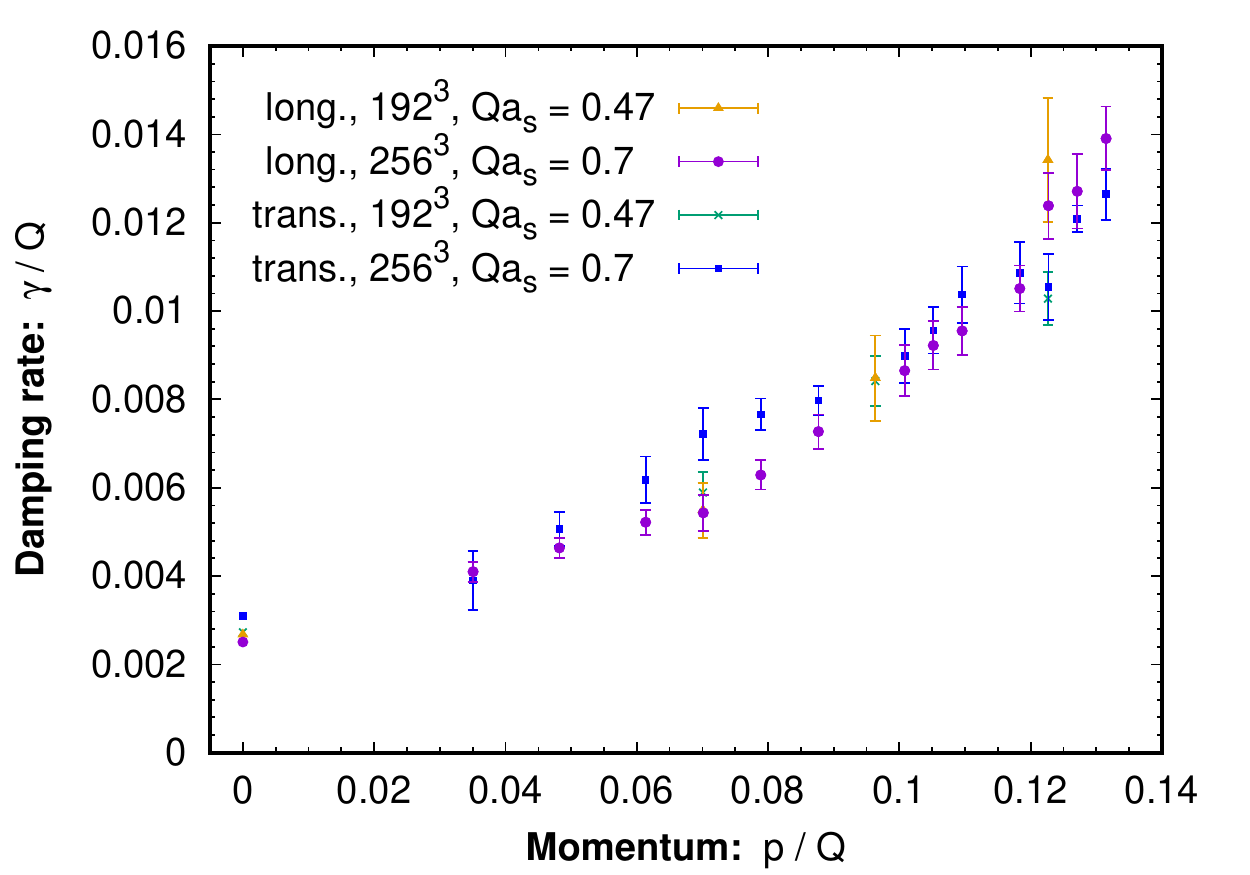}
	\caption{{\bf Upper:} Location of the maximum of $\dot{\rho}_L(\tpert,\omega,p)$ plotted logarithmically after subtracting the Landau cut region \eqref{eq_Landau_cut_long} for each momentum. For comparison, the extracted transverse dispersion $\omega_T(p)$ from \fig\ref{fig_disprel_trans} is also included. The red solid and dashed lines show the expected longitudinal and transverse dispersion relations from HTL calculations at LO, respectively (\se\ref{sec_HTL_theory}). 
	{\bf Lower:} The longitudinal and transverse damping rates (see also \fig\ref{fig_damp_rate} for the latter), extracted by fitting to a damped oscillator \eqref{eq_damped_oscil}. Error bars consist of a sum of statistical and fitting error. For the data points at $p=0$ we fitted a damped oscillator to already averaged data within $\Q \Delta t \leq 100$, thus not obtaining error bars.} 
	\label{fig_disprel_long}
\end{figure}

\subsection{Longitudinal correlations}
\label{sec_long_corrs}

Apart from the transversely polarized correlation functions, we can also study longitudinal correlation functions at unequal time. Our data is shown for the spectral function $\dot{\rho}_L$ in \fig\ref{fig_rho_dt_w_long} in relative time and frequency domains for different momentum modes, where we choose it to satisfy the sum rule \eqref{eq_rho_sum_rules}. In the time domain, one observes damped oscillations similar to what was observed for the transverse case. For larger momenta $p \gtrsim 0.15\, \Q$ the oscillations decrease very quickly and oscillations are barely visible at later times. 

Their Fourier transforms are shown in the right panel of \fig\ref{fig_rho_dt_w_long} together with the longitudinal spectral function computed within the HTL formalism at LO that was discussed in \se\ref{sec_HTL_theory}. For small momenta $p \lesssim 0.15\, \Q$, one sees clear quasiparticle peaks and a small Landau cut region, which is consistent with the HTL curve. At larger momenta $p \gtrsim 0.15\, \Q$, the HTL Landau cut region becomes more pronounced and starts to dominate the sum rule \eqref{eq_rho_sum_rules} while the residue of the quasiparticle peak gets exponentially suppressed as $~\sim \exp(-p^2/m^2)$. At the same time, the longitudinal dispersion relation of the quasiparticle peak gets exponentially close to the light cone $\omega_L \approx p$, where the Landau cut region starts. Therefore, it becomes difficult to distinguish between the Landau cut region and the quasiparticle peak numerically. Moreover, the Landau cut region gets smeared a little around the light cone, such that one sees practically only the Landau cut for $p \gtrsim 0.29\, \Q$. 

We also show the normalized longitudinal statistical correlation function $\ddot{F}_L / \ddot{F}_L(t,\Delta t = 0,p)\; \dot{\rho}_L^\HTL(t,\Delta t = 0,p)$ as orange curves in both relative time and frequency domains. This quantity is noisy but one observes an overall good agreement with the spectral function. Although it seems as for $p = 0.12\,\Q$ it does not capture the Landau cut region correctly, at larger momenta where the Landau cut region starts to dominate the spectrum, it agrees quite well with the spectral function. This confirms the existence of an approximate generalized fluctuation-dissipation relation similar to the transverse case in \eqref{eq_ddF_drho_rel} also for the longitudinal correlations. The connection between $\ddot{F}_L$ and $\dot{\rho}_L$ is provided by $\ddot{F}_L(t,\Delta t = 0,p)$ that was depicted in \fig\ref{fig_ET_vs_EL_corrs}. 

The longitudinal dispersion relation $\omega_L(p)$ can be extracted from the spectral function by, for instance, finding its maximum after subtracting the Landau cut. Since numerically, the Landau cut region appears as smeared around the light cone, while the analytic function has a steep increase there, the subtraction is not precise in the region where the Landau cut starts dominating the spectrum. Therefore, the longitudinal dispersion extracted this way is not more than an estimate for the true longitudinal dispersion relation for momenta $p \gtrsim m$. 

Our results for this are shown in the upper panel of \fig\ref{fig_disprel_long} for two different discretizations. The transverse dispersion relation $\omega_T(p)$ from \fig\ref{fig_disprel_trans} as well as the longitudinal and transverse dispersion relations from the HTL formalism at LO are added to the figure. One observes that the longitudinal dispersion estimate is quite distinct from the transverse one and exhibits very similar momentum dependence as the corresponding HTL curve. As expected, there are differences to the latter around $p \sim m_\HTL$, which may have resulted from the subtraction of the analytical Landau curve instead of the numerical one. On the other hand, at low momenta one sees a similar deviation from the HTL curve at LO as for the transverse distribution in \fig\ref{fig_disprel_trans}, where the extracted frequencies are systematically larger. This behavior is seen for both discretizations. Moreover, both dispersion relations come close, smoothly approaching $\wplas$ with the value \eqref{eq_wplas_data}, which is larger than expected from LO HTL, as discussed in \se\ref{sec_HTL_disp_damp}. Hence, $\omega_L(p)$ confirms the observations in that section. 

For momenta where the Landau cut does not dominate the spectral function, one can extract the longitudinal damping rate $\gamma_L(p)$ by fitting to a damped oscillator \eqref{eq_damped_oscil}. We show the extracted data points in the lower panel of \fig\ref{fig_disprel_long}, where we also include the transverse damping rate from \fig\ref{fig_damp_rate} for comparison. For larger momenta $p \gtrsim m_\HTL$, the Landau cut becomes important, eventually dominating the $\Delta t$ evolution of $\dot{\rho}$, and the fitting procedure provides information about the Landau cut rather than providing $\gamma_L(p)$. Therefore, the lower panel of \fig\ref{fig_disprel_long} describes only low momenta $p \lesssim m_\HTL$ correctly. For these, one observes that the damping rate does not depend on the polarization within uncertainties. A similar observation was made for the equal-time correlation function $\ddot{F}(t, \Delta t = 0, p)$ in \fig\ref{fig_ET_vs_EL_corrs}.

\begin{figure}[t]
	\centering
	\includegraphics[scale=0.72]{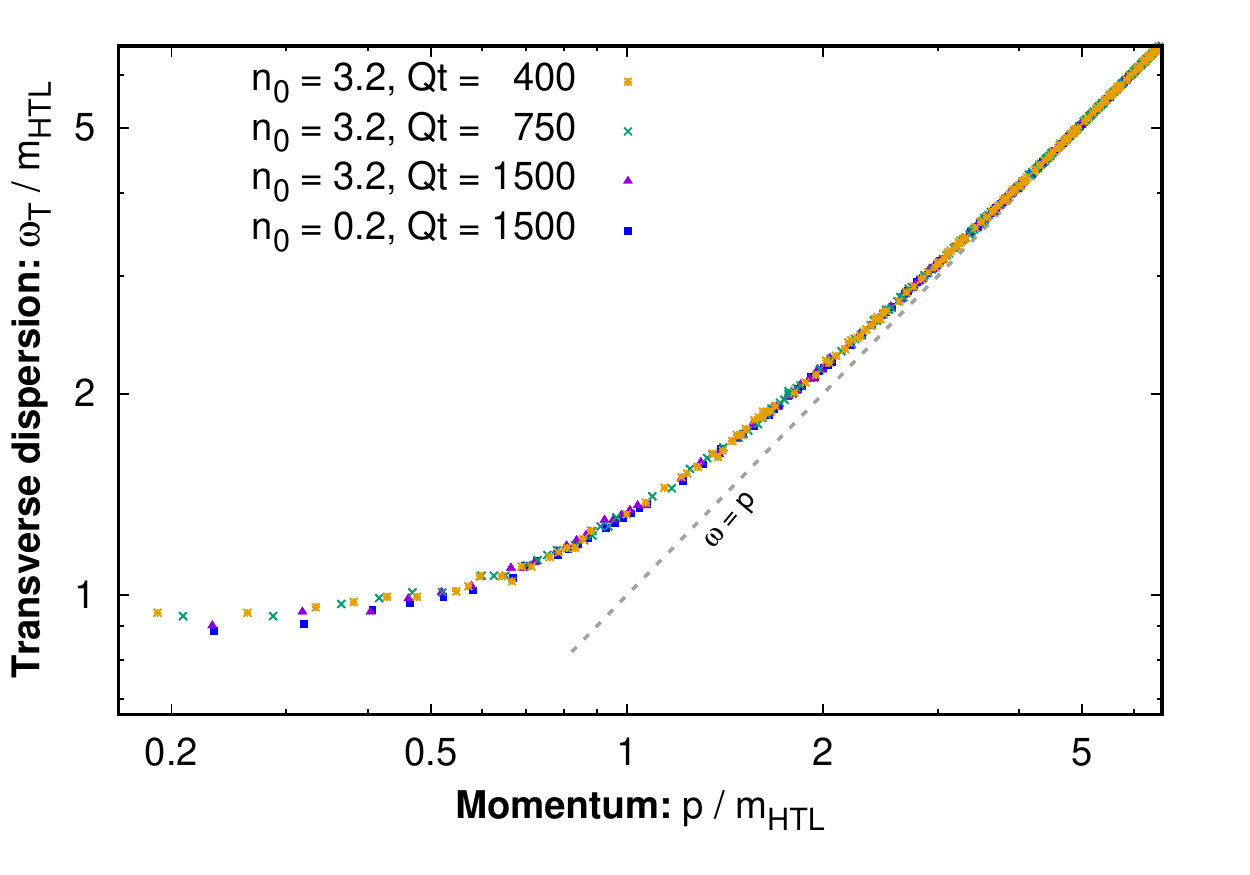}
	\caption{Transverse dispersion relation $\omega_T(p)$ for different amplitudes $n_0$ and times $\tpert$, deduced from $\dot{\rho}_T(\tpert,\omega,p)$ by finding the location of the maximum of the quasiparticle peak for each momentum, where $\Q \dtmax = 100$ was used for the Fourier transform. Both $\omega_T$ and $p$ are divided by the HTL mass $m_{\HTL}$.}
	\label{fig_disprel_trans_params}
\end{figure}

\begin{figure}[t]
	\centering
	\includegraphics[scale=0.72]{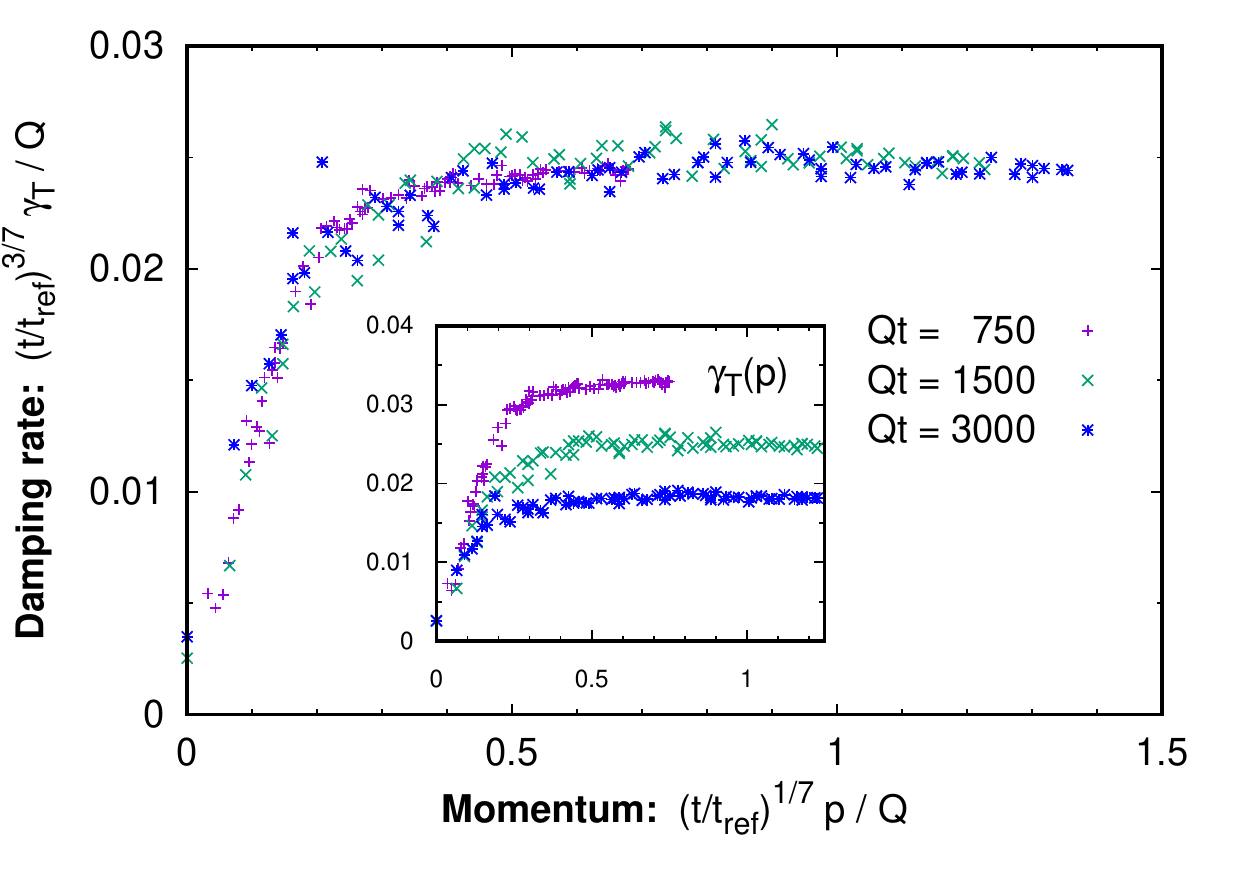}
	\caption{Transverse damping rate $\gamma_T(p)$ for different times $\tpert$ and $n_0 = 3.2$, deduced from $\dot{\rho}_T(\tpert,\Delta t, p)$ by fitting to the form of a damped oscillator \eqref{eq_damped_oscil} within a time range of $\Q \Delta t \leq 50$. In the main panel both axis are rescaled by powers of $t / t_{\mrm{ref}}$ with $\Q t_{\mrm{ref}} = 1500$ while the original data is shown in the inset without rescaling.}
	\label{fig_damp_rate_params}
\end{figure}

\subsection{Variation of parameters, approach to LO HTL}

We finally check how our results depend on a variation of the initial amplitude $n_0$ and time $\tpert$. Understanding the dependence on time is especially important in our approach. Within HTL, the coupling constant is $g \sim m/\Lambda$. Since in our case this ratio decreases with time as $m/\Lambda \sim (\Q t)^{-2/7}$, a weak-coupling limit in HTL corresponds to a late-time limit in our method. Hence, the spectral function is expected to approach HTL at LO with time. We will discuss our findings at the example of the transverse dispersion relation and damping rate. 

Our results on the dispersion relation are shown in \fig\ref{fig_disprel_trans_params} for simulations with $n_0 = 3.2$ at three different times $\Q \tpert = 400$, $750$ and $1500$ and for our $n_0 = 0.2$ curve at time $\Q \tpert = 1500$ from \fig\ref{fig_disprel_trans}. All simulations have been performed on $256^3$ lattices with $\Q a_s = 0.7$ and the $n_0 = 3.2$ simulations have not been ensemble averaged. The curves are plotted in units of $m_{\HTL}$ that is computed as in Sec.~\ref{sec_class_nonAb_theory} with $f_{\mrm{EE}}(t,p)$. While the mass is time-dependent as shown in \fig\ref{fig_msqr_diff_n0}, the dispersion relation is seen to hardly change with time. 
This is predicted by the HTL formalism, where the mass is the only parameter of the dispersion relation, and hence, expressing frequencies and momenta in terms of the mass should lead to the same form. 

We note, however, that a residual time dependence is not ruled out in our simulations. In particular, one would expect that the ratio $\wplas/m$ decreases with time, approaching the LO value $\sqrt{2/3}$ of HTL. For our lowest momenta $p \ll m_\HTL$, the frequency seems to slightly decrease with time for the $n_0 = 3.2$ dispersions and we have indeed seen a decreasing $\wplas/m$ ratio with time. However, we emphasize that the curves have not been ensemble averaged and may have a considerable uncertainty. Therefore, whether $\wplas/m$ indeed approaches $\sqrt{2/3}$ at late times cannot be answered accurately with the present data and is left for future work. 

The transverse damping rate $\gamma_T(p)$ is shown in \fig\ref{fig_damp_rate_params} for different times, rescaled by $(t / t_{\mrm{ref}})^{3/7}$ with $\Q t_{\mrm{ref}} = 1500$ as a function of rescaled momentum $(t / t_{\mrm{ref}})^{1/7} p$. The exponent $1/7$ corresponds to the evolution of the mass $1/m_\HTL$, while the exponent $3/7$ in the damping rate reflects the evolution of the effective temperature 
\begin{align}
 g^2 T_*(t)/\Q \sim \frac{\int \ud^3 p\, (g^2 f)^2}{\int \ud^3 p\; g^2 f / p\;\Q} \sim (\Q t)^{-3/7},
\end{align}
which follows from the self-similar evolution of the distribution function. We checked explicitly that $g^2 T_*(t)$ follows this power law in our simulations. Since the rescaled data coincide while the original one does not as seen in the inset, this indicates that the damping rate is proportional to $\gamma_T(p) \propto g^2 T_*$ for all momenta and not only for the zero mode as given by Eq.~\eqref{eq_gamma_0}. 
This also implies that the damping rate itself is of subleading order since $\gamma_T(p) / \Lambda \sim (m/\Lambda)^2 \ll m/\Lambda$, which is of course expected in HTL. Hence, the width of the quasiparticle peak decreases in the late-time limit and the peak approaches a Delta function, recovering the HTL prediction at LO.
However, note that all curves in \fig\ref{fig_damp_rate_params} are single-run simulations and therefore, more statistics is needed to confirm this observation beyond doubt. A precision measurement of the damping rate may also show possible influence of the magnetic scale, which is of the same subleading order $g^2 \Q$ but decreases with a different power law than $g^2 T_*$ \cite{Berges:2017igc}.


\section{Conclusion}
\label{sec:conclu}

In this work, we have studied the spectral and statistical correlation functions $\rho$ and $F$ at unequal times of highly occupied classical Yang-Mills theory at the self-similar regime to get further insight into the dynamics. To compute the spectral function, we have used linear response theory with our recently developed formalism to simulate linearized fluctuations on top of a classical background. With these techniques, we were able to access information about the highly populated system that had not been accessible previously by only considering equal-time correlation functions. Our results were compared to the HTL formalism at LO. 

We showed that the functional forms of the transverse and longitudinal spectral functions $\rho_T$ and $\rho_L$ agree well with the HTL-predicted curves, including the Landau cut region of low frequencies $\omega^2 \leq p^2$. For larger frequencies, both $\rho_T$ and $\rho_L$ involve a quasiparticle peak with a dispersion relation and damping rate. The dispersion relations agree well with the corresponding HTL curves but also with a relativistic dispersion in the transverse case while the damping rate is observed to be proportional to an effective temperature $T_*$ and to roughly agree with the predicted value in \cite{Braaten:1990it} at $p=0$. While transverse and longitudinal dispersion relations follow distinct functional forms, the corresponding damping rates are observed to agree within error bars for momenta below the mass. 

For the dispersion relations, we found some deviations at low and high momenta. Extracting the plasmon frequency $\wplas$ and asymptotic mass $m$ from our data, we were able to quantify some of these deviations. Instead of the HTL prediction $\wplas / m = \sqrt{2/3}$, we found a value close to $1$ for the ratio. Moreover, we observed that the spectral and statistical correlation functions are not connected by the effective temperature $T_*$ as predicted by the HTL formalism but by a momentum dependent function that is of the same order of magnitude for momenta $m \lesssim p \lesssim \Lambda$. 

Many general properties of the HTL correlation functions at LO in $m / \Lambda$ have been observed in our simulations. The deviations from the HTL expressions may have resulted from higher order corrections since for the studied times, we did not have a perfect scale separation between the mass and the hard scale. In our approach, $m / \Lambda$ decreases with time and therefore, our late-time limit is expected to correspond to the LO of the HTL framework. We have not performed this limit systematically in this work, but we have seen indications of this for instance in the transverse damping rate, which indeed appears to be of subleading order $\gamma_T(p) / m \sim m / \Lambda$. Therefore, the quasiparticle peaks should approach Delta functions for $m / \Lambda \rightarrow 0$, as expected from HTL at LO. We have not gone to large enough times with sufficient statistics to see the approach to LO HTL also in the dispersion relation. A more systematic study of the time evolution of the spectral function is left for future work. 

One of the achievements of the new method is that we were able to measure the transverse and longitudinal damping rates for the first time, extending the result from \cite{Braaten:1990it} to finite momenta. In this sense, our simulations can also be regarded as a non-perturbative way of measuring observables that are hard to obtain analytically within the HTL formalism. We plan to use this numerical method also for similar cases, where HTL quantities are hard to access. 

Moreover, in the future we aim to extend our studies to the case of a background field with a very anisotropic momentum distribution. Here the fluctuations are known to be unstable, i.e., one has to map out both the real and imaginary parts of the dispersion relation, and with different angles of orientation between the momentum vector and the anisotropy. The advantage of the formulation in terms of linearized fluctuations is that this method does not rely on a scale separation between the hard and soft modes, although we have here operated in a regime where such a separation exists. Thus, for the isotropic case, the calculation can be extended to small values of $\Q t$ before the onset of the self-similar regime. More importantly for the phenomenological context, the calculation can be performed in a longitudinally expanding coordinate system and extended all the way to $\tau=0$. This will enable one to follow the growth of the quantum fluctuations in a realistic geometry corresponding to a relativistic heavy ion collision. We intend to pursue these avenues in future work.


\begin{acknowledgments}
  We are grateful to J\"{u}rgen Berges, Jacopo Ghiglieri and Asier Pi\~{n}eiro Orioli for valuable discussions and would like to thank Asier Pi\~{n}eiro Orioli for sharing with us the method in Eq.~\eqref{eq_source_relations} in a private communication. T.~L.\ is supported by the Academy of Finland, projects No. 267321 and No. 303756. This work is supported  by the European Research Council, grant ERC-2015-CoG-681707. J.~P.\ is supported by the Jenny and Antti Wihuri Foundation. J.~P.\ acknowledges support for travel from Magnus Ehrnrooth foundation. K.~B.\ and J.~P.\ would like to thank CERN and its Theory group for hospitality during part of this work. The authors wish to acknowledge CSC – IT Center for Science, Finland, for computational resources. 
\end{acknowledgments}


\appendix

\begin{figure}[t]
	\centering
	\includegraphics[scale=0.7]{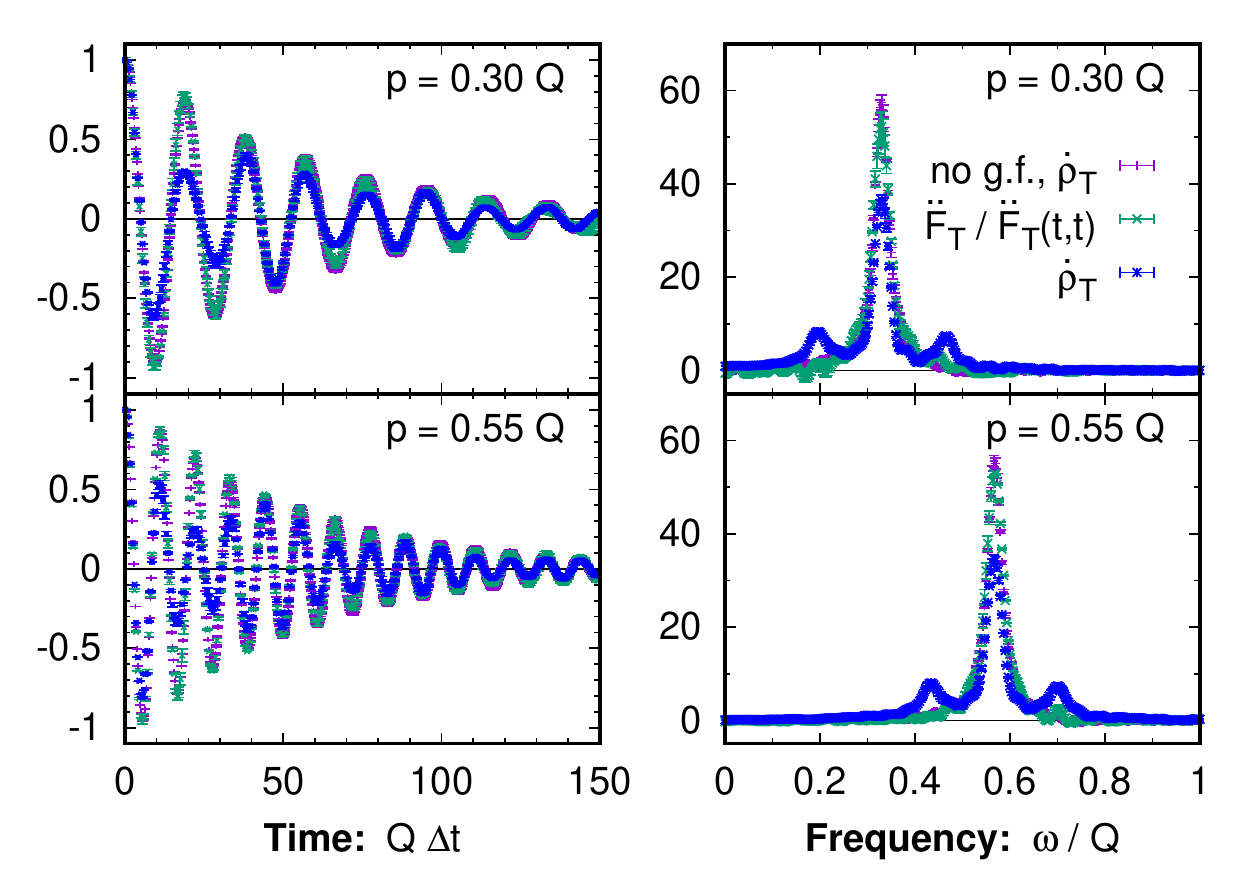}
	\caption{Shown are the spectral and statistical correlation functions with subsequent gauge fixing in both temporal and frequency domain. For comparison, the spectral function from \figs\ref{fig_rho_F_dt_multi} and \ref{fig_rho_F_w_multi} is shown, where no subsequent gauge fixing is involved (abbreviated by `no g.f.').}
	\label{fig_comp_gf}
\end{figure}

\section{Correlation functions with subsequent gauge fixing}
\label{app:gaugefix}

In this appendix we discuss the correlation functions for the case when after the introduction of the linearized fluctuations, the background field is gauge fixed to Coulomb gauge every time before correlation functions are printed. They are shown for a $256^3$ lattice with $\Q a_s = 0.7$ in \fig\ref{fig_comp_gf} in both temporal and frequency domains, where we averaged over two simulations. While the peak is at the same position as the spectral function computed as discussed in the main sections, one finds that the temporal evolution introduces some spurious amplitude modulations. These translate into side peaks in frequency space. Interestingly, these artifacts are absent in the statistical correlation function $\ddot{F}$, which therefore follows a Lorentzian form and matches with our usual computation of the spectral function. This implies that both the dispersion relation as well as the damping rate of the peaks in $\ddot{F}$ are close to those measured above. The amplitude modulations and thus, the side peaks of the spectral function are quite sensitive to the lattice spacing and the volume, showing clearer side peaks and stronger modulations with increasing volume.


\bibliographystyle{JHEP-2modlong}
\bibliography{spires}

\end{document}